\documentclass[10pt,journal,twoside]{IEEEtran}

\usepackage[autostyle=true]{csquotes} 
\usepackage{graphicx}
\usepackage{epstopdf}
\usepackage{amsmath}
\usepackage{amssymb}
\usepackage{amsthm}
\usepackage{mathtools}
\usepackage{bbm}
\usepackage{color}
\usepackage{booktabs}
\usepackage{blkarray}
\usepackage{subfig}
\usepackage{cite}
\usepackage[justification=raggedright,singlelinecheck=false]{caption}

\newtheorem{theorem}{Theorem}

\newcommand{\vx}{\ensuremath{\mathbf x}}
\newcommand{\vy}{\ensuremath{\mathbf y}}

\newcommand{\vu}{\ensuremath{\mathbf u}}
\newcommand{\vw}{\ensuremath{\mathbf w}}
\newcommand{\vs}{\ensuremath{\mathbf s}}
\newcommand{\vv}{\ensuremath{\mathbf v}}
\newcommand{\ve}{\ensuremath{\mathbf e}}
\newcommand{\vq}{\ensuremath{\mathbf q}}
\newcommand{\va}{\ensuremath{\mathbf a}}

\newcommand{\vxh}{\widehat{\vx}}

\newcommand{\vuh}{\widehat{\vu}}
\newcommand{\vyh}{\widehat{\vy}}
\newcommand{\vqh}{\widehat{\vq}}
\newcommand{\vqo}{\overline{\vq}}
\newcommand{\vqt}{\ensuremath{\widetilde{\vq}}}
\newcommand{\vsh}{\widehat{\vs}}
\newcommand{\vxt}{\widetilde{\vx}}
\newcommand{\vyt}{\widetilde{\vy}}
\newcommand{\vyth}{\ensuremath{\widehat{\vyt}}}
\newcommand{\why}{\ensuremath{\widehat{y}}}
\newcommand{\whq}{\ensuremath{\widehat{q}}}

\newcommand{\wtY}{\ensuremath{\widetilde{Y}}}
\newcommand{\FF}{\ensuremath{\mathbb{F}}}

\newcommand{\sC}{\ensuremath{\mathcal{C}}}

\newcommand{\mA}{\ensuremath{\mathbf A}}

\newcommand{\mH}{\ensuremath{\mathbf H}}
\newcommand\refX[2]{\ensuremath{X_{0_{#1, #2}}}}
\newcommand{\Reals}{\ensuremath{\mathbb{R}}}
\newcommand{\Integers}{\ensuremath{\mathbb{Z}}}

\newcommand\inner[2]{\langle #1, #2 \rangle}

\newcommand{\E}{\mathbb{E}}
\newcommand{\Var}{\mathrm{Var}}

\newcommand{\sT}{\ensuremath{\mathcal{T}}}
\newcommand{\sI}{\ensuremath{\mathcal{I}}}

\begin{document}

\title{Distributed Coding of \\Quantized Random Projections}

\author{Maxim Goukhshtein, Petros T. Boufounos, Toshiaki Koike-Akino, and Stark
C. Draper
\thanks{Manuscript received August, 2019; revised July, 2020.}
\thanks{Maxim Goukhshtein and Stark C. Draper are with the department of
Electrical and Computer Engineering, University of Toronto, Toronto, ON,
Canada, email: {\em maxim.goukhshtein@mail.utoronto.ca},
{\em stark.draper@utoronto.ca}}
\thanks{Petros T. Boufounos and Toshiaki Koike-Akino are with Mitsubishi
Electric Research Laboratories (MERL), Cambridge, MA 02139, USA, email: {\em
$\lbrace$petrosb, koike$\rbrace$@merl.com}}
\thanks{MG and SD are partially supported by the National Science Foundation
	(NSF) under Grant CCF-1217058, and by the Natural Science and Engineering
	Research Council (NSERC) of Canada, including through a Discovery Research
	Grant. MG has also been partially supported by the Queen Elizabeth II
	Graduate Scholarship, the Ontario Graduate Scholarship and the NSERC
	Postgraduate Scholarship. MG did part of this work while at MERL. PG and TK
	are exclusively supported by MERL. Portions of this material were presented
	at the 2017 IEEE Int. Symp. Info. Theory, Aachen, Germany, Jun. 2017.}}

\markboth{IEEE Transactions on Signal Processing,~Vol.~XX, No.~X, Xxxxxx~2020}%
{Goukhshtein \MakeLowercase{\textit{et al.}}: Distributed Coding of Quantized Random Projections}

\maketitle

\begin{abstract}
    In this paper we propose a new framework for distributed source coding of
    structured sources, such as sparse signals. Our framework capitalizes on recent
    advances in the theory of linear inverse problems and signal representations
    using incoherent projections. Our approach acquires and quantizes incoherent
    linear measurements of the signal, which are represented as separate bitplanes.
    Each bitplane is coded using a distributed source code of the appropriate rate,
    and transmitted. The decoder, starts from the least significant biplane and,
    using a prediction of the signal as side information, iteratively recovers each
    bitplane based on the source prediction and the signal, assuming all the
    previous bitplanes of lower significance have already been recovered. We
    provide theoretical results guiding the rate selection, relying only on the
    least squares prediction error of the source. This is in contrast to existing
    approaches which rely on difficult-to-estimate information-theoretic metrics to
    set the rate. We validate our approach using simulations on remote-sensing
    multispectral images, comparing them with existing approaches of similar
    complexity.
\end{abstract}

\begin{IEEEkeywords}
	Distributed source coding, lossy compression, quantization, sparsity, compressed sensing, syndrome decoding, low complexity encoder, side information.
\end{IEEEkeywords}

\section{Introduction}
\label{sec:intro}
The increasing availability of data, due to the growth of sensing applications,
has made compression indispensable in modern signal processing systems. Most
modern approaches, such as JPEG, JPEG-2000 and HEVC, rely on some form of
transform coding to exploit the signal structure, often after signal prediction
is preformed at the encoder. These approaches typically exhibit higher
computational complexity at the encoder, opting for a simpler decoder. However,
in some applications, such as remote sensing, it is necessary to have
lightweight encoders, and shift the complexity to the decoder.

One solution to this problem has been Distributed Source Coding (DSC), first
introduced in~\cite{SlepianWolf} for lossless compression of discrete sources,
and extended in~\cite{WynerZiv} for lossy compression of continuous sources.
Several practical approaches exploit DSC in the context of
image~\cite{Schonberg1,Schonberg3} and video data~\cite{Girod05,TotoZarasoa12},
as well as remote sensing data~\cite{Rane,Wang}, among others. A common
drawback of these approaches is that the encoder often has to be designed and
tuned specifically for the application, in order to be able to exploit the
structure of the source.

In addition to the design difficulties, one of the key issues in deploying
practical DSC systems is the rate control at the encoder. In particular, DSC
relies on side information available at the decoder to assist decoding of the
compressed source. The choice of an appropriate compression rate, determined at
the encoder as a function of the quality of side information available at the
decoder, is critical for the success of these methods; at high compression
rates, the decoder might not have sufficient side information to decode the
source at all.

To control the rate, DSC literature typically relies on information-theoretic
metrics, such as mutual information. Unfortunately, for many real-world
sources, these can be difficult to quantify, especially using a lightweight
encoder. Thus, a number of practical systems either reduce their compression
rate to guarantee that the signal can be decoded, or rely on feedback to inform
the encoder that the information received is sufficient for decoding.
Nevertheless, both options have drawbacks: the former reduces compression
efficiency and the later requires an active bidirectional connection between
the encoder and the decoder. Rate-control methods using easier-to-quantify
metrics, such as the mean squared prediction error, would make rate estimation
easier at the encoder.

This paper introduces a new approach to lossy distributed compression, assuming
a prediction of the source signal is available at the decoder. Our approach
capitalizes on recent advances in linear inverse problems, to apply DSC to
quantized linear measurements. Specifically, we rely on efficiently obtaining
quantized linear measurements of the source at the encoder, separating
measurements to bitplanes and coding each bitplane using syndrome-based DSC.
The decoder exploits the prediction and the syndromes to iteratively predict
bitplanes and decode the quantized measurements. Once the quantized
measurements are recovered, the decoder solves an inverse problem to recover
the signal, taking its structure into account.

Our approach has three distinct advantages:
\begin{enumerate}
      \item {\em Universality:} in contrast to most approaches, the encoder
            design requires no knowledge of the source structure. Only the
            decoder uses the source structure, during reconstruction and,
            possibly, in forming the prediction.
      \item {\em Simple Rate Control:} the rate required to transmit syndromes
            can be explicitly computed based on an upper bound of the $\ell_2$
            error of the decoder's signal prediction. This is much easier to
            estimate, and more readily available at the encoder, than
            information-theoretic measures used in existing DSC-based approaches.
      \item {\em Low Encoding Complexity:} even compared to DSC-based schemes,
            the proposed encoding method is very lightweight and straightforward
            to implement. Complexity is dominated by a matrix-vector
            multiplication, which can exploit efficient fast transforms such as
            the Hadamard transform.
\end{enumerate}

To demonstrate the applicability of our approach, we provide an example on
multispectral satellite image compression. This work significantly improves
on~\cite{Valsesia} by introducing a DSC framework to encode the signal. A
preliminary version of this work, focused on the multispectral image
compression application, appeared in~\cite{Maxim}. In this paper we
significantly expand on~\cite{Maxim} by (a) introducing new tools that improve
the decoding of DSC syndromes and incorporate likelihood estimates in the
belief propagation, and (b) introducing a successive decoding approach, in
which the decoding of each spectral band is exploited to improve the side
information and the prediction at the decoder for decoding the next band.

The next section establishes notation and provides a brief background on
distributed source coding and linear inverse problems.
Section~\ref{sec:methodology} presents the core framework of our approach.
Section~\ref{sec:implementations} discusses practical aspects for efficient
implementation. Section~\ref{sec:results} discusses the application to
multispectral image compression and presents our experimental results.
Section~\ref{sec:conclusion} provides a brief discussion and concludes.
\section{Background}
\label{sec:background}

\subsection{Distributed Source Coding}
\label{subsection:DSC}
Distributed source coding, introduced for lossless compression by Slepian and
Wolf~\cite{SlepianWolf}, allows compression of a source by an encoder, given
side information which is only available at the decoder, to be made as
efficient as if the side information was also available at the encoder. More
generally, the Slepian-Wolf theorem states that lossless compression, with
separate encoders and a joint decoder, of two sources $ U $ and $ V $ drawn
from $p(u,v)$, can be achieved with any rate satisfying
\begin{align*}
    R_{U} \geq H(U|V),~R_{V} \geq H(V|U),~R_{U} + R_{V} & \geq H(U,V),\label{eq:X_SW_bnd}
\end{align*}
where $ H(\cdot|\cdot) $ and $ H(\cdot,\cdot) $ denote conditional and joint
entropies, respectively. An extension of the results to lossy compression was
later developed by Wyner and Ziv~\cite{WynerZiv}. A comprehensive treatment of
DSC can be found in~\cite[Ch.~15]{CoverThomas}
and~\cite[Ch.~10-12]{ElGamalKim}.

The connection between distributed source coding and channel coding also has
roots in the same period~\cite{Wyner,Ancheta}. Nonetheless, it was only about
30 years later that a practical approach to perform DSC using syndrome decoding
of channel codes was developed by Pradhan and Ramchandran \cite{DISCUS}.

In this approach, the side information at the receiver generates a prediction
of the source. Since the prediction contains errors, the encoder needs to
transmit sufficient information to correct these errors. The prediction is
treated as the output of a noisy channel, introducing the errors. Thus, the
transmitter can use a channel code to provide redundancy and correct the errors
introduced by the channel/prediction. The channel code is designed to use a
parity check matrix that computes a syndrome comprising of parity check bits
that provide sufficient redundancy to correct the prediction errors. The rate
of the syndrome, i.e., the number of bits required to correct the errors,
depends on the quality of the prediction, which characterizes the capacity of
the implied channel.

In particular, assuming a binary source $ \vu\in \FF_{2}^{n} $ and its
prediction $\vv \in \FF_{2}^{n}$, the prediction can be expressed as an
unknown additive error $\ve\in \FF_{2}^{n}$ to the source, $\vv = \vu +
    \ve$. Assuming the coefficients of $ \ve $ are distributed as i.i.d
Bernoulli-$p$ random variables, where $p$ is known both at the encoder and
decoder, the decoder needs sufficient error correcting information to infer
$\ve$ and correct the prediction \vv\ to \vu. In a channel coding context,
\vv\ can be considered as the output of a binary symmetric channel (BSC)
with crossover probability $ p $ (BSC-($p$)) and input \vu.

To correct the channel effect, and recover \vu\ from \vv, we can use a
parity-check matrix $\mH \in \FF_{2}^{(n-k)\times n}$ for a code with rate $R$
and consider $ \vs_{\vu} = \mH\vu $ and $ \vs_{\vv} = \mH\vv $, the syndromes
of $ \vu $ and $ \vu $, respectively. The sum of those syndromes, equal to
\begin{equation}
    \vs_{\vu} + \vs_{\vv} = \mH(\vu + \vv) = \mH(\vu + \vu + \ve) = \mH\ve = \vs_{\ve},
\end{equation}
is the syndrome of the error pattern $ \ve $. The decoder can use the syndrome
$\vs_{\ve}$ to determine the most likely error pattern $ \widehat{\ve} $ and
estimate the original source $ \vu $ as $ \vuh = \vv+ \widehat{\ve}$. The
compression is attained by representing $ \vu \in \FF_{2}^{n} $ via its
syndrome $ \vs_{\vu} \in \FF_{2}^{n-k} $. The compression ratio $ \rho $,
defined as
\begin{equation}
    \rho = \frac{\text{\# bits to represent } \vu}{\text{\# bits to represent } \vs_{\vu}} = \frac{n}{n-k} = \frac{1}{1 - \frac{k}{n}} = \frac{1}{1 - R},
\end{equation}
is maximized when $R$ is as large as possible while still allowing for
decoding. Assuming the elements of $ \vu $ are distributed uniformly, the code
can be designed with rate up to the channel capacity of the BSC-$(p)$, i.e.,
$R_{\mathrm{max}} = C_{\mathrm{BSC}-(p)} = 1 -
    H_\mathrm{B}(p)$~\cite{CoverThomas}, where $ H_\mathrm{B}(p) =
    -p\log_{2}(p)-(1-p)\log_{2}(1-p)$ is the binary entropy function. In that case,
$ H_\mathrm{B}(p) = H(V|U) = H(U|V) $ and the highest possible compression
ratio is
\begin{equation}
    \rho_{\mathrm{max}} = 1/H_\mathrm{B}(p) = 1/H(U|V).
\end{equation}
This implies that $ \vu $ can be encoded at a rate $ R_{U} = \frac{1}{\rho_\mathrm{max}} = H(U|V) $, achieving the Slepian-Wolf bound.

\subsection{Compressed Sensing and Linear Inverse Problems}
\label{sec:background_cs_inverse_problems}
The introduction of Compressive Sensing~\cite{CandesTao1} invigorated interest
in sparse signal models, sampling, and linear inverse problems. A typical
linear inverse problem aims to recover an unknown signal $\vx\in\Reals^n$ from
measurements $\vy\in \Reals^m$, acquired using an, often underdetermined,
linear measurement matrix $\mA\in\Reals^{m\times n}$. Measurements might also
be distorted and noisy. Most relevant to this work are quantized formulations,
possibly with additive dither $\vw\in\Reals^m$, i.e., of the form
\begin{equation} \label{eq:uls}
    \vq=Q(\vy) = Q(\mA\vx+\vw)
\end{equation}
where $Q(\cdot)$ is a scalar quantizer. Signal recovery is typically formulated
as a convex optimization problem,
\begin{equation} \label{eq:opt_general}
    \widehat{\vx}=\arg\min_{\vx}~\mathcal{D}\left(\vq,
    \mA\vx+\vw\right) +\lambda \mathcal{R}(\vx),
\end{equation}
where $\mathcal{D}(\cdot,\cdot)$ is a data fidelity penalty, matching the
measurements of the reconstructed signal with the acquired quantized
measurements, and $ \mathcal{R}(\cdot) $ is a regularization function,
promoting a signal model, such as sparsity. Adherence to data fidelity is
traded off with the enforcement of the signal model through the choice of
$\lambda$. Reconstruction using~\eqref{eq:opt_general} can be efficiently
solved using, for example, proximal gradient methods~\cite{ProxAlgoritms}.

The role of the regularizer is to promote solutions that exhibit properties
consistent with the signal model of $ \vx $, e.g., sparsity or low total
variation. Often the regularizer may exploit additional prior information, such
as the existence of a signal that shares a similar sparsity pattern as \vx,
which can be incorporated by weighting the regularization
function~\cite{weightedl1}.

The role of the data fidelity term $\mathcal{D}(\cdot,\cdot)$ is to promote
adherence to the measurements. In particular for quantized measurements,
several data fidelity penalties have been proposed, mostly trying to enforce
consistent reconstruction, i.e., the measurements of the original and
reconstructed signals quantize to the same
values~\cite{Jacques2,Dai,boufounos2015quantization}. Alternatively, a simple
$\ell_2^2$ penalty, i.e., $\mathcal{D}\left(\vq, \mA\vx+\vw\right)=\left\|\vq-
    \mA\vx-\vw\right\|_2^2$, may work as well or better, especially for fine
quantization. While optimal measurement quantizers can be designed with a
variety of criteria~\cite{Sun,Kipnis}

\subsection{Source Coding and Compressive Sensing}
The intersection of quantization and compressive sensing as a source coding
approach has been studied extensively in the
literature~\cite{Boufounos1,boufounos2015quantization,Dai,Sun,Zymnis,Dai2,Bek1,Jacques,Kipnis},
including the development of optimal quantizer designs~\cite{Sun,Kipnis} and
reconstruction methods~\cite{Dai2,Bek1,Jacques}. A survey of results, both
theoretical and practical, can be found in~\cite{boufounos2015quantization}. It
is by now well established that scalar quantization is suboptimal in terms of
rate-distortion trade-off, compared to transform coding~\cite{Boufounos1}.

In particular, compressed sensing methods use a linear measurement process to
acquire signals that lie in unions of lower-dimensional subspaces.
Consequently, the resulting measurements are redundant. This redundancy can be
exploited, e.g., to provide robustness to erasures~\cite{LBDB_ACHA11}. However,
it also diminishes the rate-distortion performance of quantized compressed
sensing, even when using optimal quantizer designs. The shortcoming is inherent
in the measurement process, not in the scalar nature of the quantizer. Thus, it
is relatively straightforward to extend the results in~\cite{Boufounos1} to
demonstrate that more general union of subspace models, such as the ones
introduced in~\cite{Baraniuk10}, would also exhibit suboptimal rate-distortion
trade-offs. So would common approaches to vector quantization, such as lattice
quantizers~\cite{Gersho91}.

Ideas from distributed source coding have also made their way in the
compressive sensing literature. In
particular,~\cite{baron2009distributed,Duarte13} introduced the notion of
distributed Compressive Sensing, in which multiple sensors are used to acquire
different signals that exhibit some similarity in their structure. By combining
all the measurements and jointly reconstructing the signals it is possible to
exploit the common structure and reduce the required number of measurements per
signal, compared to the measurements required if each signal was independently
acquired and reconstructed, agnostic of the others. The resulting bounds
demonstrate behavior analogous to classical DSC behavior. However, the
framework is continuous and not immediately applicable as a source coding
approach. Direct quantization of distributed compressive measurements,
including optimal designs, have been proposed in a number of contexts,
e.g.,~\cite{Kang09,Liu10,Shirazinia14ICASSP,shirazinia2014distributed}. Still,
such schemes suffer from the same limitations as conventional compressive
sensing: the measurements are redundant and direct measurement quantization is
suboptimal in a rate-distortion sense.

In particular, after scalar quantization, the most significant bits (MSBs) of
quantized compressive measurements contain more redundant information than the
least significant bits. The redundancy is exaggerated if side information or a
good prediction of the signal already exists. This observation, first made
in~\cite{universal}, is developed and exploited
in~\cite{B_SampTA11,Valsesia,VB_ITW16,Maxim} to reduce the coding rate by
eliminating the redundancy in the MSBs. The framework we introduce in this
paper significantly improves on these results by theoretically characterizing
the redundancy of the MSBs in the presence of a prediction and using
distributed coding to eliminate it. We should note that, while our approach
significantly improves the rate-distortion trade-off, we make no claims of
rate-distortion optimality.

To improve the rate of distributed compressive measurements, distributed source
coding has also been proposed in~\cite{Magli}, with some theoretical analysis
in~\cite{coluccia2015compressed}. This approach requires very specific
probabilistic source models, similar to~\cite{baron2009distributed}, to be able
to perform reconstruction. In addition, there is no analysis of how the
syndrome rate should be determined, effectively relying on feedback to the
transmitter. Similarly
to~\cite{baron2009distributed,weightedl1,Valsesia,VB_ITW16,Maxim} and our
proposed approach,~\cite{Magli} further exploits the side information to
improve the performance of the reconstruction algorithm. The option of
quantizing the source before measuring is also considered and, as expected by
the theory, performs worse. Another approach is~\cite{Elzanaty}, which requires
coarse support estimation at the encoder and syndrome-base coding of the
coefficients in the support, thus increasing the complexity of the encoder.
Again, the approach requires a sparse signal model, making it impractical for
approximately-sparse signals or for signals exhibiting different or additional
structure, such as low total variation (TV) or signals on a
manifold~\cite{EFTEKHARI201567,weightedl1,van2018compressed,Baraniuk10}.

\section{Methodology}
\label{sec:methodology}
Our goal is to compress, i.e., encode, an arbitrary source vector $ \vx \in
	\Reals^{n} $ using a low-complexity encoder, under the assumption that $ \vxh
$, a prediction of $ \vx $, is only available to the decoder and that the
prediction error $ \epsilon = \|\vx - \vxh \|_{2} $, or an upper bound, is
known by the encoder. Fig.~\ref{fig:end_to_end_diag} shows an end-to-end
diagram of our approach, which we describe below.

\subsection{Encoding}
\begin{figure*}[ht]
	\centering
	\includegraphics[width=.9\linewidth]{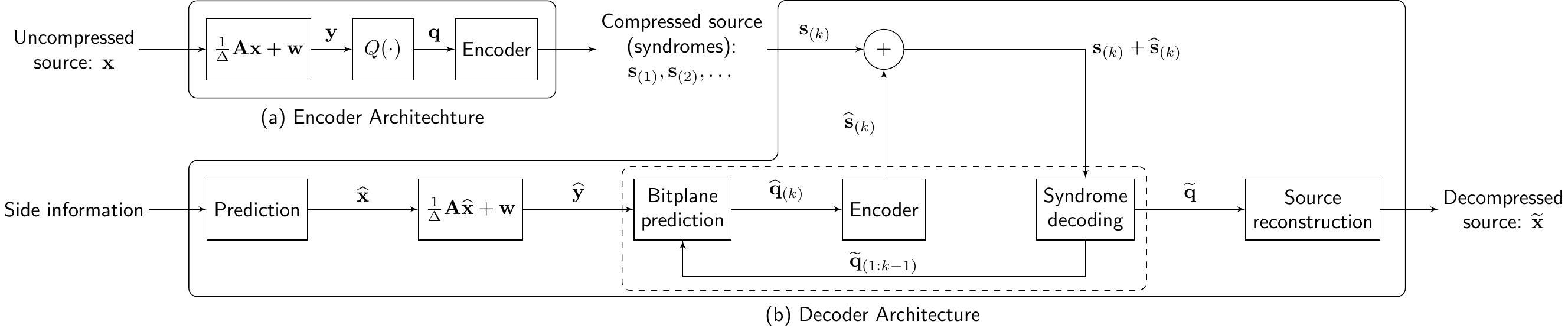}
	\caption{End-to-end high-level architecture and system diagram
		for (a) the compressor and (b) the decompressor.}
	\label{fig:end_to_end_diag}
\end{figure*}

The encoding process, shown in
Fig.~\ref{fig:end_to_end_diag}(a), consists of:
\begin{itemize}
	\item Linear measurement and quantization of the source.
	\item Syndrome generation from the quantized measurements.
\end{itemize}

\subsubsection{Measurement and Quantization}
\label{section:quant_CS_measurement}
In the first stage, the encoder generates measurements $\vy \in\Reals^m$
according to
\begin{equation}
	\vy = \frac{1}{\Delta}\mA\vx + \vw,
	\label{eq:measurement}
\end{equation}
where $\mA \in \Reals^{m\times n} $ is a measurement operator, $ \Delta \in
	\Reals^{+} $ is a scaling parameter, and the vector $ \vw \in \Reals^m $ is
a randomized dither with i.i.d. elements drawn uniformly in $[-1,0)$. The
measurement parameters, $\Delta$, \mA, and \vw, are also available at the
decoder.

The acquired measurements are quantized element-wise, using a $ B $-bit scalar
uniform integer quantizer $ Q(\cdot) $. Quantization produces the quantized
measurements $ \vq \in \Integers^m $, with the $ i^\mathrm{th} $ quantized
measurement given by
\begin{equation}
	\vq_{i} = Q(\vy_{i}) = \left\lfloor \vy_{i} + \frac{1}{2} \right\rfloor.
	\label{eq:quantization}
\end{equation}
We assume that $ B $ is selected sufficiently large, such that the quantizer does not saturate.

Since the quantizer rounds to the nearest integer, the scaling parameter
$\Delta$ is equivalent to an effective quantization interval on unscaled
measurements. Thus, the choice of $ \Delta $ affects the reconstruction quality
and the bit budget needed to encode $ \vx $. In particular, reducing $ \Delta $
results in finer measurement quantization and, thus, improved reconstruction,
at the cost of higher encoding rate.  Note that particular rate points can be
chosen simply by looking at the data and choosing $\Delta$ at the encoder,
without a need for decoding. Furthermore, the random dither $ \vw $ ensures
that the quantization error is uniformly distributed, i.e., $ \left(Q(\vy_{i})
	- \vy_{i}\right) \stackrel{i.i.d}{\sim} \mathcal{U}[-\frac{1}{2},\frac{1}{2}) $
for $ i = 1, \dots, m $ \cite{Schuchman}, a property that facilitates the
analysis necessary to determine the appropriate rate to use in encoding each
syndrome.

\subsubsection{Syndrome Generation} \label{section:syndrome_generation} In
order to encode and generate syndromes for the quantized measurements, we first
separate them into distinct bitplanes. In particular, we use $ \vq_{(k)} $, $
	k=1, \ldots, B $, to denote the $ k^\mathrm{th} $ bitplane of the quantized
measurements. We use the convention that $ k = 1 $ and $ k = B $ respectively
represent the least significant bit (LSB) plane and the most significant bit
(MSB) plane. The binary representation of $ \vq $ can be thought of as an $ m
	\times B $ binary matrix whose $ k^\mathrm{th} $ column, $\vq_{(k)}\in \FF_2^m,
$ is a binary vector containing the $ k^\mathrm{th} $ significant bit of all $
	m $ quantized measurements of $ \vx $. The following is an example of the above
described bitplane representation with $ m = 3 $ and $ B = 4 $:

\begin{equation*}
	\vq = \left(\begin{array}{c} 3  \\  0  \\ 1 \end{array}\right) \overset{\FF_{2}}{\Longrightarrow}
	\begin{blockarray}{cccc}
		\vq_{(4)} & \vq_{(3)} & \vq_{(2)} & \vq_{(1)} \\
		\begin{block}{(cccc)}
			0  & 0  & 1 & 1 \\
			0  & 0  & 0 & 0 \\
			0  & 0  & 0 & 1 \\
		\end{block}
	\end{blockarray}.
\end{equation*}

For each bitplane $ \vq_{(k)} $ the encoder generates a syndrome at a specific
rate. The rate is different for each bitplane and is based on the expected
number of errors in this bitplane that will need to be corrected. The syndromes
are used by the decoder to correct these errors and reconstruct $ \vx $ via the
iterative decoding procedure outlined in Section~\ref{sec:decoding}.

In particular, for each bitplane, $\vq_{(k)}$, the encoder assumes that the
decoder has an estimate of the bitplane, $\vqh_{(k)}$ with a few errors, i.e.,
bitflips. This estimate is computed at the decoder using the side information
and the previously corrected bitplanes $1,\ldots,k-1$. The encoder assumes that
each bit in this estimate might be flipped with probability $p_{k}$, which is
assumed known at the encoder. Thus, as described in Sec.~\ref{subsection:DSC},
the prediction acts as a binary symmetric channel with crossover probability
$p_{k}$. The encoder can efficiently encode $\vq_{(k)}$ by transmitting a
syndrome that corrects the bitflips in the predicted bitplane, i.e., using a
code with rate lower than:
\begin{equation}
	C_{\mathrm{BSC}-(p_{k})} = 1 - H_\mathrm{B}(p_{k}).
	\label{eq:bsc_capacity}
\end{equation}
Thus, the encoder uses a parity-check matrix $ \mH_{(k)} $ with associated rate
lower than~\eqref{eq:bsc_capacity}, to allow correction of bitplane prediction
errors. In other words, the encoder computes syndromes $\vs_{(k)} =
	\mH_{(k)}\vq_{(k)} $  for $ k = 1, \dots, B$ and transmits them to the decoder.
In practice, as elaborated in Section~\ref{sec:cutoff_pr}, only few lower
significance bitplanes need to be encoded; higher significance ones can be
perfectly predicted at the receiver.

Of course, to be able to successfully recover $ \vq_{(k)} $, it is critical
that the encoder accurately determines the probabilities $ p_{k} $, or an upper
bound for them. The following theorem makes this computation exact, as a
function of the $\ell_2$ norm of the prediction error, assuming that the
measurement matrix \mA\ is drawn from an i.i.d. Gaussian distribution.

\begin{theorem}
	\label{thm:flip_prob}
	Consider a signal $ \vx $ with measurements $\vy$ acquired using
	\eqref{eq:measurement} and quantized to $\vq$ using
	\eqref{eq:quantization}, where $\mA$ is randomly drawn with i.i.d.
	$\mathcal{N}(0,\sigma^2)$ entries. Assume there exists a prediction $ \vxh
	$, with prediction error $\epsilon = \|\vx - \vxh \|_{2}$ and that the
	first $k-1$ least significant bitplanes of $\vq$, namely  $ \vq_{(i)}, i =
		1, \ldots, k-1 $, are perfectly known. Then $ \vq_{(k)} $, the $
		k^\mathrm{th} $ bitplane of $\vq$, can be estimated from $ \vxh $ and $
		\vq_{(i)}, i = 1, \ldots, k-1 $ with bit error probability equal to
	\begin{align}
		p_k & = \frac{1}{2}-\sum_{l=1}^{+\infty}
		e^{-\frac{1}{2}\left(\frac{\pi\sigma\epsilon
					l}{2^{k-1}\Delta}\right)^2}\mathrm{sinc}\left(\frac{l}{2^k}\right)\mathrm{sinc}\left(\frac{l}{2}\right).
		\label{eq:bit_error}
	\end{align}
\end{theorem}

\begin{IEEEproof} See Appendix \ref{appx:thm1_prf}.
\end{IEEEproof}

As detailed in the proof, this error probability can be achieved using the
decoding method described in Sec.~\ref{sec:decoding} below.
Fig.~\ref{fig:probability_curves}(a) plots the theoretical and empirical bit
error probabilities for bit $ k = 3$. The theoretical values are calculated
using Thm.~\ref{thm:flip_prob} as a function of the prediction error $ \epsilon
	= \|\vx - \vxh \|_{2} $. The empirical values are calculated as $
	m^{-1}\|\vq_{(3)} - \vqh_{(3)}\|_{0} $, where $ \vq_{(3)} $ and $ \vqh_{(3)} $
correspond to the third bitplane of the quantized measurements of $ \vx $ and $
	\vxh $, respectively, generated from synthetic data such that $ \epsilon =
	\|\vx - \vxh \|_{2} $. As evident in the figure, the empirical probabilities
closely match the ones predicted by Thm.~\ref{thm:flip_prob}.

Note that, even if the prediction error is not known but can be upper bounded,
the theorem provides an upper bound on the probability of bitflips.
Furthermore, as we discuss in Sec.~\ref{sec:implementations}, the theorem can
be used to provide guidance even if other, more practical, measurement matrices
are used.

\begin{figure*}[!ht]
	\centering
	(a)\includegraphics[width=.3\linewidth]{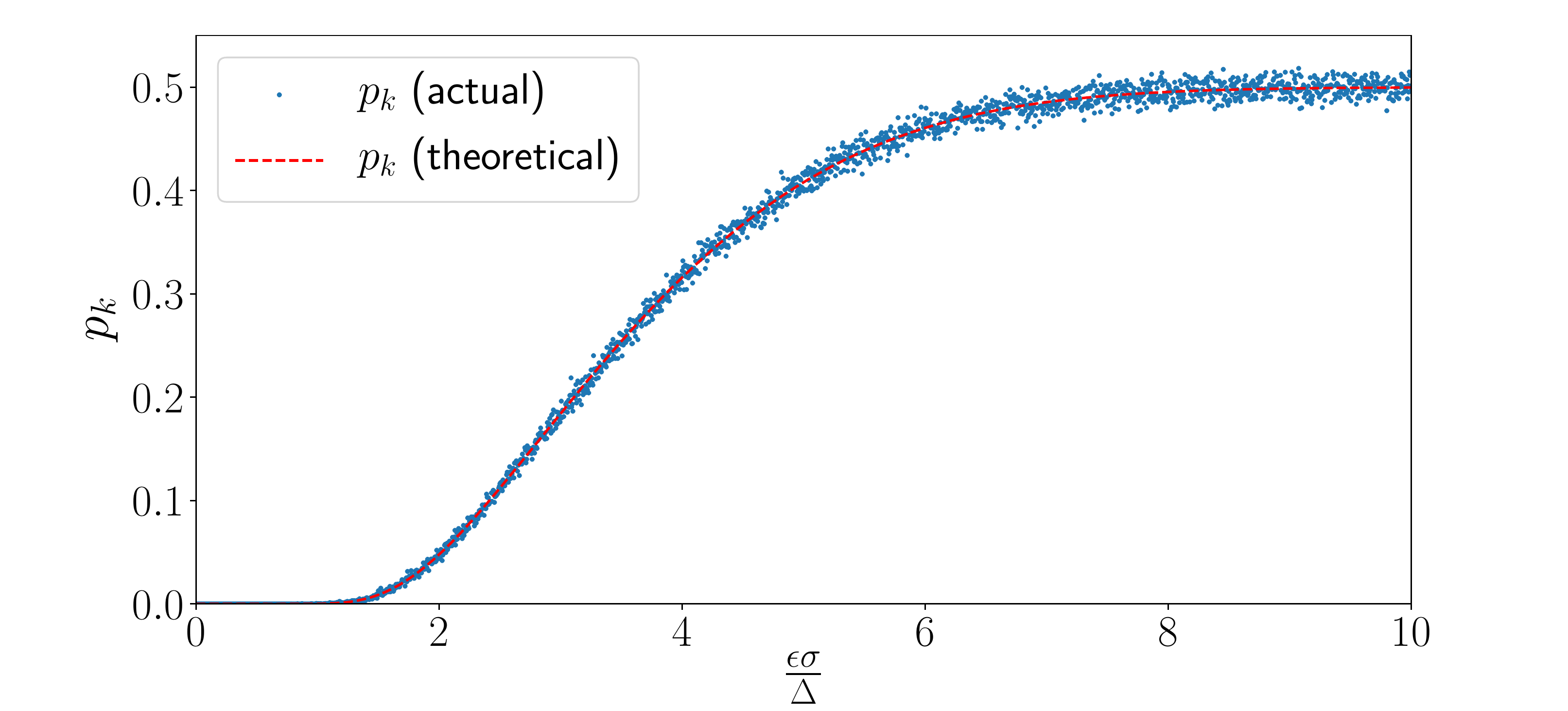}
	(b)\includegraphics[width=.3\linewidth]{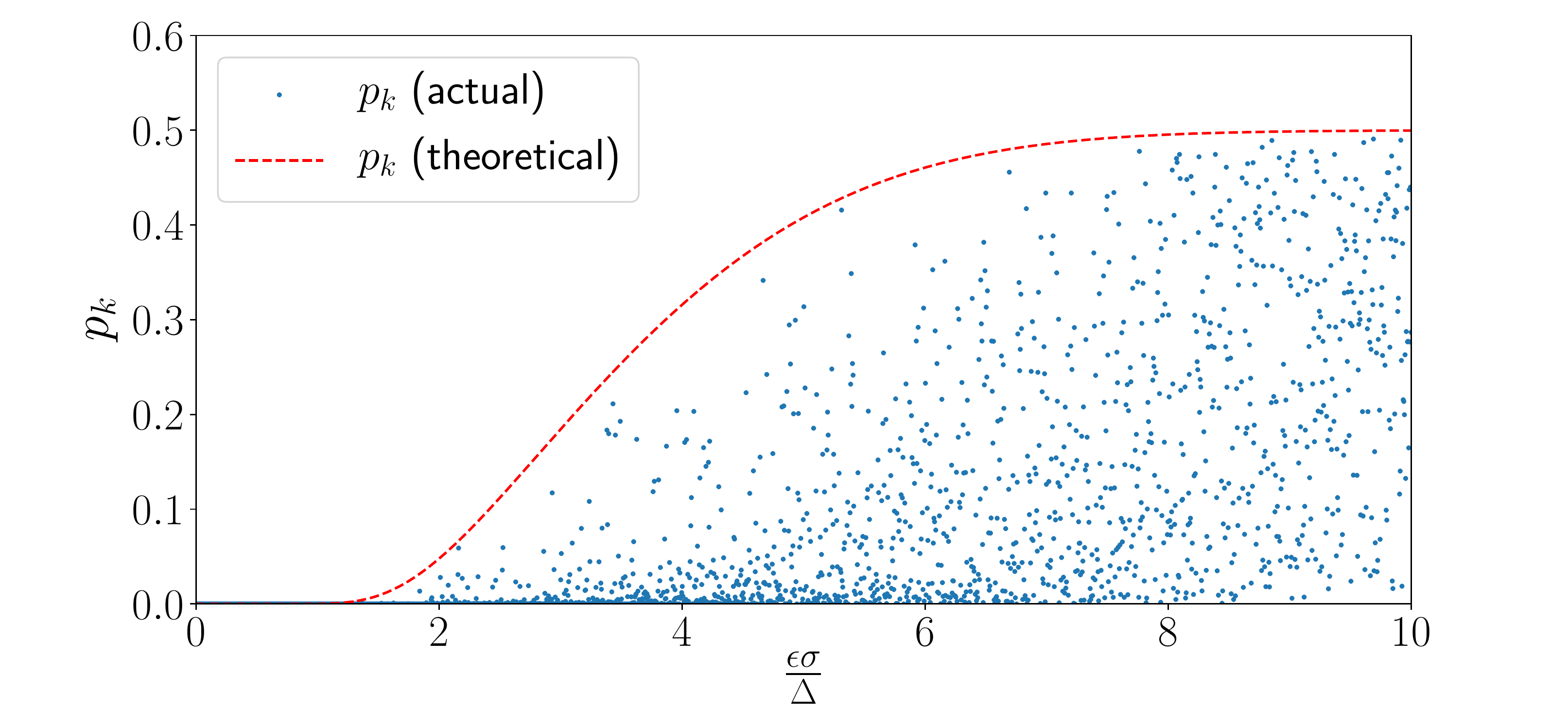} (c)
	\includegraphics[width=.3\linewidth]{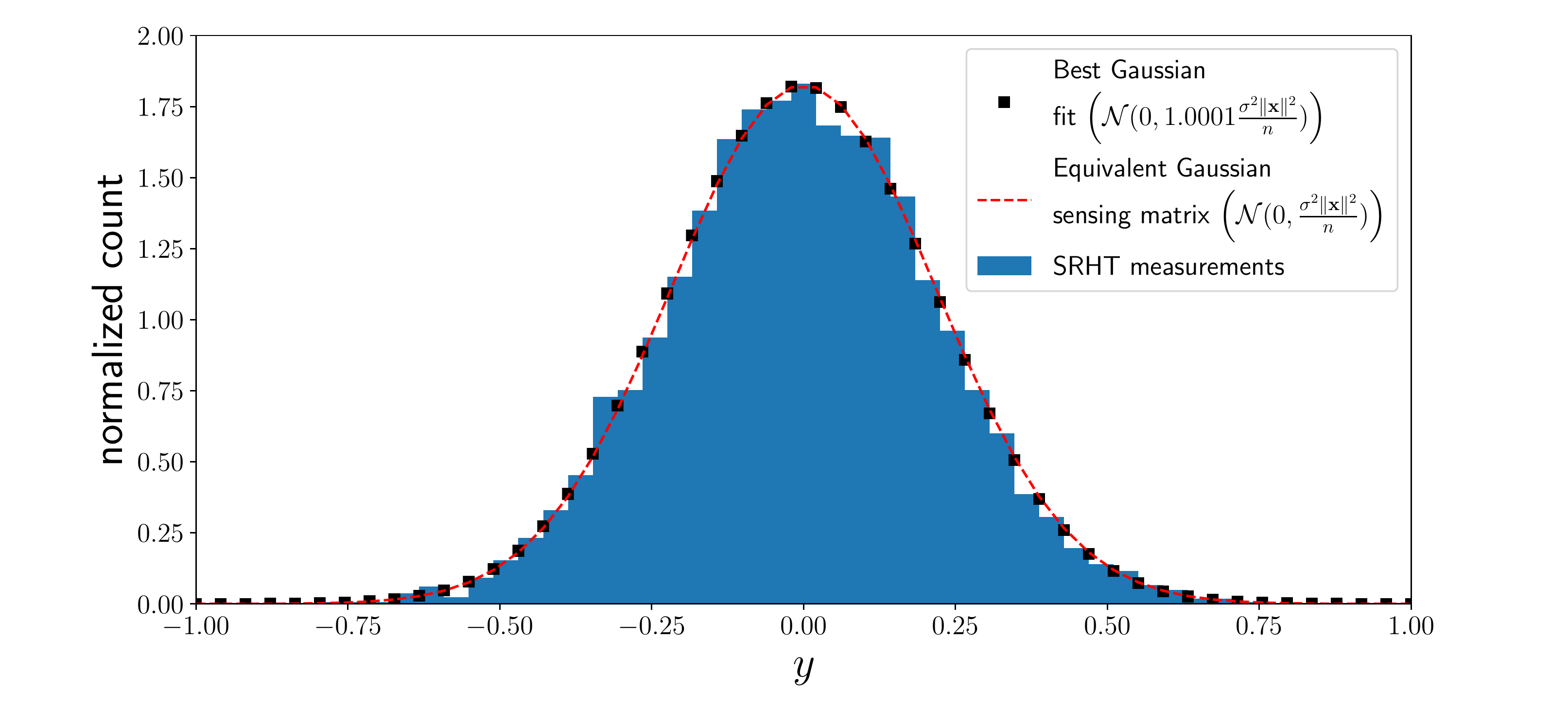}
	\caption{Empirical and theoretical bit error probabilities for quantized
		measurements as a function of the normalized prediction error $
			\frac{\epsilon\sigma}{\Delta} $ for $ k = 3 $, (a) acquired using a
		Gaussian matrix, and (b) acquired using a SRHT operator. (c) Histogram of
		measurements acquired using the SRHT operator, a Gaussian fit using the
		empirical mean and variance of the measurements and the equivalent Gaussian
		sensing matrix statistics.}

	\label{fig:probability_curves}
\end{figure*}

\subsection{Decoding}
\label{sec:decoding}
The decoder, shown in Fig.~\ref{fig:end_to_end_diag}(b), uses the syndromes $
	\vs_{(k)} $ of the quantized measurement bitplanes $ \vq_{(k)} $ and the
side information to recover the quantized measurements and reconstruct the
signal. Decoding, or decompression, consists of:
\begin{itemize}
	\item Measurement prediction from the side information.
	\item Recovery of the original quantized measurements via iterative bitplane prediction and syndrome decoding.
	\item Source reconstruction from the quantized measurements.
\end{itemize}

\subsubsection{Prediction and Measurement}
First, the decoder generates a prediction $ \vxh $ of the source $ \vx $ using
the side information and measures the prediction using the same measurement
parameters $ \mA, \Delta $, and $ \vw$ as used by the encoder:
\begin{equation}
	\vyh = \frac{1}{\Delta}\mA\vxh + \vw. \label{eq:pred_meas}
\end{equation}

\subsubsection{Recovery of Quantized Measurements} \label{Sec:quant_meas_recov}
\begin{figure}[t!]
	\centering
	\includegraphics[width=.7\linewidth]{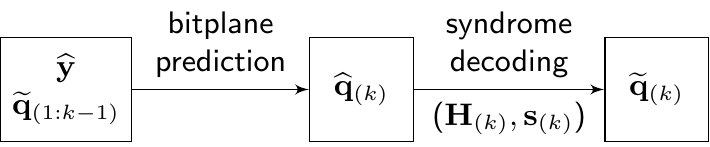}
	\caption{Recovering quantized measurements of bitplane $ k $\textbf{} from the predicted measurements $ \vyh $ and the previously recovered $ k - 1 $ bits $ \vqt_{(1:k-1)} $.}
	\label{FIG:pred_meas_flow}
\end{figure}

Next, the decoder reconstructs the original quantized measurements $\vq$ one
bitplane at a time, starting from the least significant bitplane. For the
recovery of each bitplane, the decoder uses all the previously recovered
bitplanes and the predicted measurements $ \vyh $. As illustrated in
Fig.~\ref{FIG:pred_meas_flow}, each syndrome-encoded bitplane $ k $ is
recovered via a two-stage scheme: bitplane prediction followed by syndrome
decoding.

The quantized measurements are iteratively recovered starting with the least
significant bitplane $k=1$.  At iteration $k$, a new estimate of the quantized
measurements $ \vqh $ is computed using bitplane prediction, incorporating all
the already decoded information from previous iterations. From that estimate,
the $k^\mathrm{th}$ bitplane, $\vqh_{(k)}$, is predicted and then corrected
using the syndrome $\vs_{(k)}$, to recover the corrected bitplane $\vqt_{(k)}$.
If the syndrome rate has been correctly chosen, decoding is successful with
high probability and $ \vqt_{(k)}=\vq_{(k)}$.

Specifically for $k=1$, the decoder estimates the quantized measurements $\vqo
	= Q(\vyh)$ and extracts the least significant bitplane $\vqh_{(1)}=\vqo_{(1)}$.
The corrected least significant bitplane, $\vqt_{(1)}$, is obtained by
correcting the mismatch between $\vqh_{(1)}$ and $\vq_{(1)}$ using the
syndromes $\vs_{(1)}$ and $ \vsh_{(1)} = \mH_{(1)}\vqh_{(1)} $. For the
remaining bitplanes, $k>1$, assuming $k-1$ bitplanes have already been
successfully decoded, $\vqh$ is estimated by selecting the uniform quantization
interval consistent with the decoded $k-1$ bitplanes and closest to the
prediction $\vyh$. Having correctly decoded the first $k-1$ bitplanes is
equivalent to the signal being encoded with a $(k-1)$-bit universal quantizer
\cite{universal}. Thus, recovering $\vqt $ uses the same decoding as
in~\cite{Valsesia}.

\begin{figure}[t]
	\centering
	\includegraphics[width=0.5\textwidth]{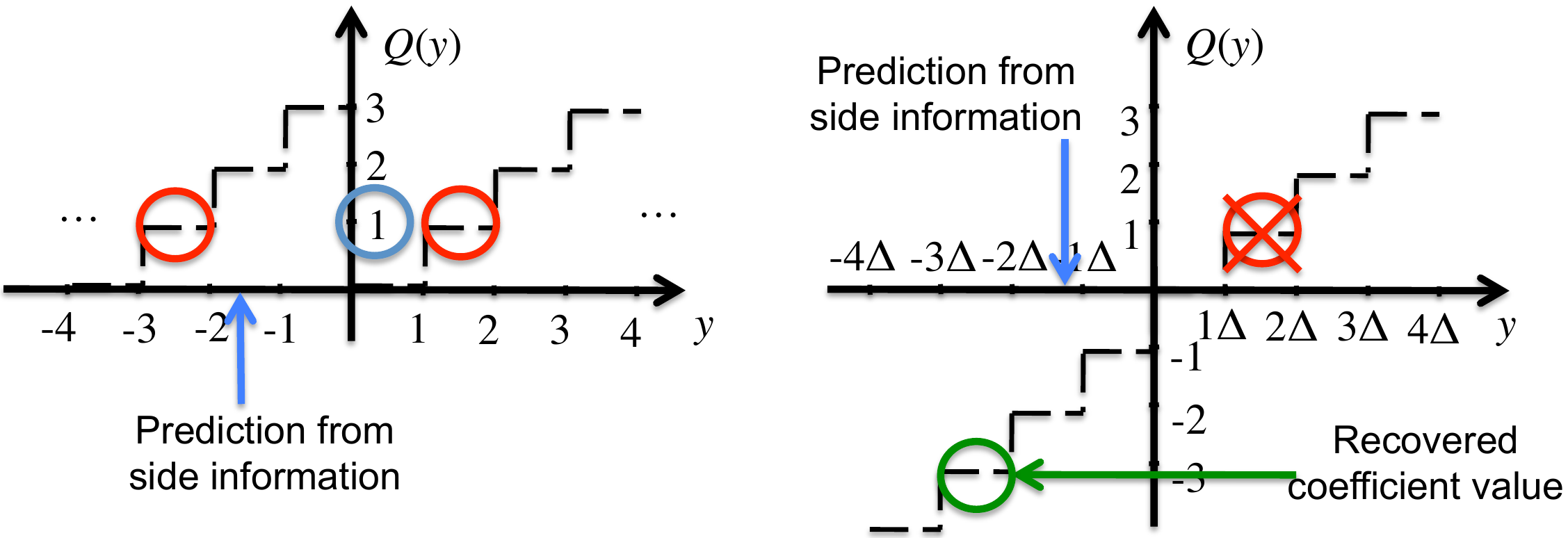}
	\caption{Bitplane prediction of a quantization point using the prediction measurement.}
	\label{FIG:pred_res}
\end{figure}

An example of $k-1=2$ is shown in Fig.~\ref{FIG:pred_res}. The left hand side
of the figure plots a 2-bit universal quantizer, equivalent to a uniform scalar
quantizer with all but the 2 least significant bits dropped. The right hand
side shows the corresponding 3-bit uniform quantizer used to produce $ q $. In
this example, the two least significant bits decode to the universal
quantization value of 1, which could correspond to $ q = 1 $ or $-3$ in the
uniform quantizer. However, the prediction of the measurement $ \why $ is
closer to the interval corresponding to $ q = -3 $, and, therefore $ \whq = -3
$ is recovered. For a formal description of the bitplane prediction process see
Appendix~\ref{appx:thm1_prf}.

After the $k^\mathrm{th}$ bitplane is estimated, the estimate $ \vqh_{(k)}$ is
corrected via syndrome decoding of $\vs_{(k)}$ and $\vsh_{(k)}$ to produce the corrected estimate
$ \vqt_{(k)}$. As long as the syndromes satisfy the rate conditions of
Thm.~\ref{thm:flip_prob}, decoding is reliable and $ \vqt_{(k)}=\vq_{(k)}$.

Decoding continues iteratively until all $B$ bitplanes have been correctly
decoded. Reliable decoding at every iteration guarantees that the decoded
quantized measurements are equal to the encoded quantized measurements, i.e., $
	\vqt=\vq$. The described syndrome coding procedure bears conceptual similarity
to multilevel coding schemes where information is encoded using a number of
channel codes of different rates and decoding proceeds in a multistage fashion
where each bitplane, starting from the LSB, is decoded by incorporating
information from preceding stages~\cite{Imai}. We  note that, although in
theory the decoding order of the bitplanes should not matter---as long as the
encoding provides the correct syndrome rates given this order, and the decoder
is able to exploit all statistical dependencies---decoding from least to most
significant bit can reduce the number of bitplanes that need to be coded. In
practice, this reduces the rate overhead due to finite block lengths and other
practical considerations.

We should also note that a common assumption in such schemes is that decoding
is successful in each stage before decoding the next. Incorrect decoding of one
stage will degrade the bitplane prediction for the next stage and affect the
probability of correct decoding. Since the probability of incorrect decoding
can, in principle, become arbitrarily low using good codes of the correct
rates, the probability of failing to decode any of the $B$ bitplanes can be
easily controlled using a union bound on the probability of failure to decode
each bitplane separately. A careful analysis of the effects of incorrect
prediction on the error probability at the next bit level could be performed
using the tools used in proving Theorem~\ref{thm:flip_prob}. However, such an
analysis is beyond the scope of this work.

\subsubsection{Source Reconstruction}
The decoder solves an inverse problem to reconstruct the source from the
recovered quantized measurements $\vqt=\vq$, as described in
Sec.~\ref{sec:background_cs_inverse_problems}:
\begin{equation} \label{eq:opt_recon}
	\vxt=\arg\min_{\vx}~\mathcal{D}\left(\widetilde{\vq},
	\frac{1}{\Delta}\mA\vx+\vw\right) +\lambda \mathcal{R}(\vx).
\end{equation}
The regularizer $\mathcal{R}(\cdot)$ may exploit the side information, such as
the prediction's sparsity pattern, to improve the signal model.

\section{Implementation Considerations}
\label{sec:implementations}

\subsection{Efficient Measurements Operators}
Using a Gaussian matrix for encoding, as assumed in Thm.~\ref{thm:flip_prob},
may prove challenging particularly when the encoder is implemented in
finite-precision arithmetic with limited memory and computation. Instead, most
practical implementations will use measurement approaches based on fast
transforms~\cite{WHT_RIP}, some of which have been shown to satisfy the
restricted isometry property (RIP) or have other properties desirable for
measurement systems. These approaches exploit the fast transform structure to
reduce computation and storage to $\mathcal{O}(n\log n)$ and $\mathcal{O}(n)$,
respectively, instead of $\mathcal{O}(nm)$.

Approaches based on the Walsh--Hadamard transform (WHT) are particularly
appealing, as the transform only uses addition and subtraction operations, and
no multiplications. Similarly to the approach used in~\cite{Duarte08}, we
measure by applying a random permutation to the signal $ \vx $, applying the
WHT and randomly subsampling the output. We denote this SRHT, although it is
slightly different than the SRHT in~\cite{Tropp}, in that our approach does not
multiply the signal with a random diagonal $\pm 1$ operator before the
permutation.

The histogram in Fig.~\ref{fig:probability_curves}(c) illustrates that our SRHT
measurements behave similarly to measurements of the same signal using a
Gaussian matrix that satisfies Thm.~\ref{thm:flip_prob} conditions.
Fig.~\ref{fig:probability_curves}(b) further suggests empirically that the bit
error probability computed in Thm.~\ref{thm:flip_prob} overestimates $p_k$,
when the SRHT is used---we conjecture it is an upper bound. Thus, it seems
reasonable to use~\eqref{eq:bit_error} in this case, something we confirmed in
our experiments in Section~\ref{sec:results}.

\subsection{Binary Representation}
To represent the quantized measurements, we use an offset-binary
representation. In other words, all measurements are made positive before
quantization by an appropriate constant shift that can be removed in decoding.
This representation produces quantization intervals that are uniform in all
bitplanes. Furthermore, combined with the randomized measurements and our
choice of dither, this representation produces uniform $\{0,1\}$ bit
distribution in all bitplanes, which is important in designing the syndromes. A
two's complement binary representation has similar properties and could be used
instead.

In contrast, sign-magnitude and one's complement representations have two
different binary strings representing zero; a positive and a negative one.
These require special handling, and make the representation more cumbersome and
slightly less efficient. In addition, such representations might introduce bias
and correlations on the bit distribution in each bitplane.

\subsection{Efficient Channel Codes}\label{section:LDPC_likelihood} In order to
maximize the syndrome-based compression, a rate-efficient channel code must be
used to generate syndromes. One popular choice are low-density parity-check
(LDPC) codes which are capacity-approaching, and hence allow high compression
rates. LDPC codes are particularly appealing because their extremely sparse
parity-check matrices allow syndrome generation with complexity $ \sim
	\mathcal{O}(m)$, rather than $ \mathcal{O}(m^{2})$. Furthermore, the sparsity
enables simple efficient decoding using belief propagation.

Decoding using belief propagation is initialized using prior likelihoods
associated with each of the bits to be decoded, reflecting the prior
probability of a certain bit taking the value 0 versus 1. In our case, the
likelihood corresponds to the probability that the predicted bit in question
will be in error (when compared to the same bit of the measured signal). A
simple agnostic option is to set the likelihood for all the bits of a given
bitplane $ k $ to its associated bitflip probability $ p_{k} $. Since the bits
in each bitplane are uniformly distributed by design, as described above, it is
not necessary to adjust this prior, as described, for example,
in~\cite{Chen09a,Chen09b}.

In addition, the prediction process in Fig.~\ref{FIG:pred_res} provides more
information that can be used to improve this likelihood estimate individually,
for each bit in the bitplane, using the predicted measurements, $ \vyh $,
according to the following theorem.

\begin{theorem}
	\label{thm:likelihood}
	Consider a signal $ \vx $ that is measured and quantized in the same manner
	as described in Thm.~\ref{thm:flip_prob}. Assume there exists a prediction
	$ \vxh $, with prediction error $\epsilon = \|\vx - \vxh \|_{2}$, and let $
		y $ and $ \why $ denote single measurements of $ \vx $ and $ \vxh $,
	respectively. Assume also that the first $k-1$ least significant bits of
	$Q(y)$, namely $ q_{(i)}, i=1,\ldots,k-1 $, are perfectly known. Then, the
	likelihood of error in $ \whq_{(k)}$, the $ k^\mathrm{th} $ bit of $
		Q(\why)$, can be estimated from $ \why $ and $ q_{(i)}, i=1,\ldots,k-1 $, as
	\begin{align}
		L_{k} & = \mathrm{Pr}\left(\whq_{(k)} \neq q_{(k)} \mid \why, q_{(1)},\ldots,q_{(k-1)}\right) \\
		      & = \frac{A_{2}(k,c)}{A_{1}(k,c) + A_{2}(k,c)},
		\label{eq:bit_likelihood}
	\end{align}
	where
	\begin{align}
		 & A_{1}(k, c) \nonumber                                                                                                                                                                                               \\
		 & = \frac{1}{2^{k}}\left(1 + 2 \sum_{l=1}^{+\infty} e^{-\frac{1}{2}\left(\frac{\pi\sigma\epsilon l}{2^{k-1}\Delta}\right)^2} \cos\left(\frac{\pi cl}{2^{k-1}}\right)\mathrm{sinc}\left(\frac{l}{2^{k}}\right)\right), \\
		 & A_{2}(k,c) = A_{1}(k, 2^{k-1} - c),
	\end{align}
	and where $ c $ is the smallest distance from $ \why $ to the center of the
	quantization interval consistent with the $ k - 1 $ LSBs, $ q_{(i)},
		i=1,\ldots,k-1 $.
\end{theorem}

\begin{IEEEproof}See Appendix \ref{appx:thm2_prf}. \end{IEEEproof}

Fig.~\ref{fig:reliability_behavior}(a) demonstrates how the error likelihood $
	L_{k} $ behaves as a function of the distance parameter $ c $, for $ k = 3 $
and different values of the normalized prediction error $
	\frac{\epsilon\sigma}{\Delta} $. As evident, the error likelihood maximum value
of 0.5 is attained when $ c = \frac{2^{3}}{2} = 4 $, namely when the prediction
$ \why $ falls exactly in the middle between the two possible consistent
quantization intervals (i.e., exactly in between the closest $ A_{1} $ and $
	A_{2} $ regions, in which case $ c = \frac{2^{k}}{2} = 2^{k-1} $). In that
case, the uncertainty in the $ k^\mathrm{th} $ bit's predicted value, is
maximized, resulting in the maximum likelihood of error.

\begin{figure*}[]
	\centering
	(a)\includegraphics[width=.3\linewidth]{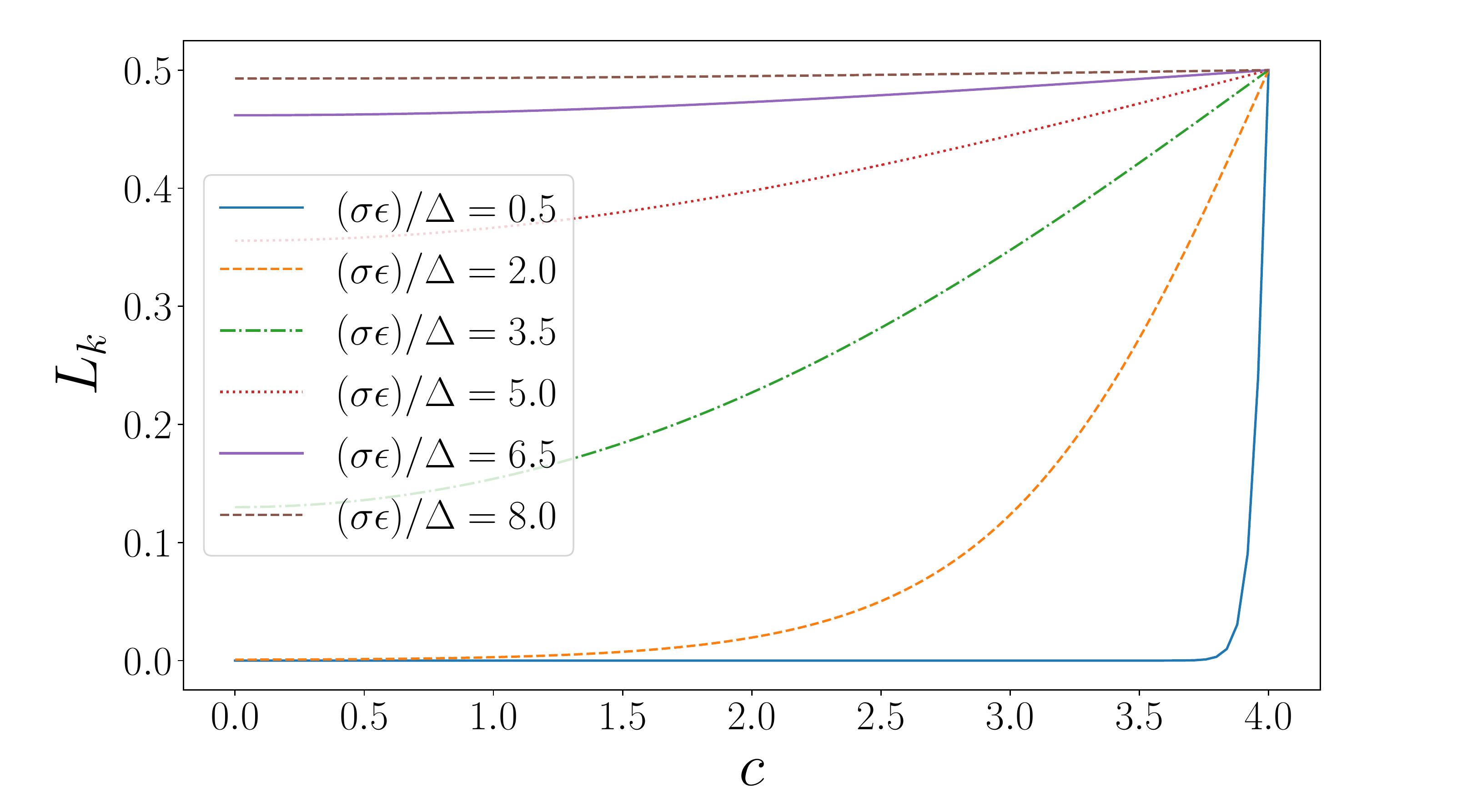}
	(b)\includegraphics[width=.3\linewidth]{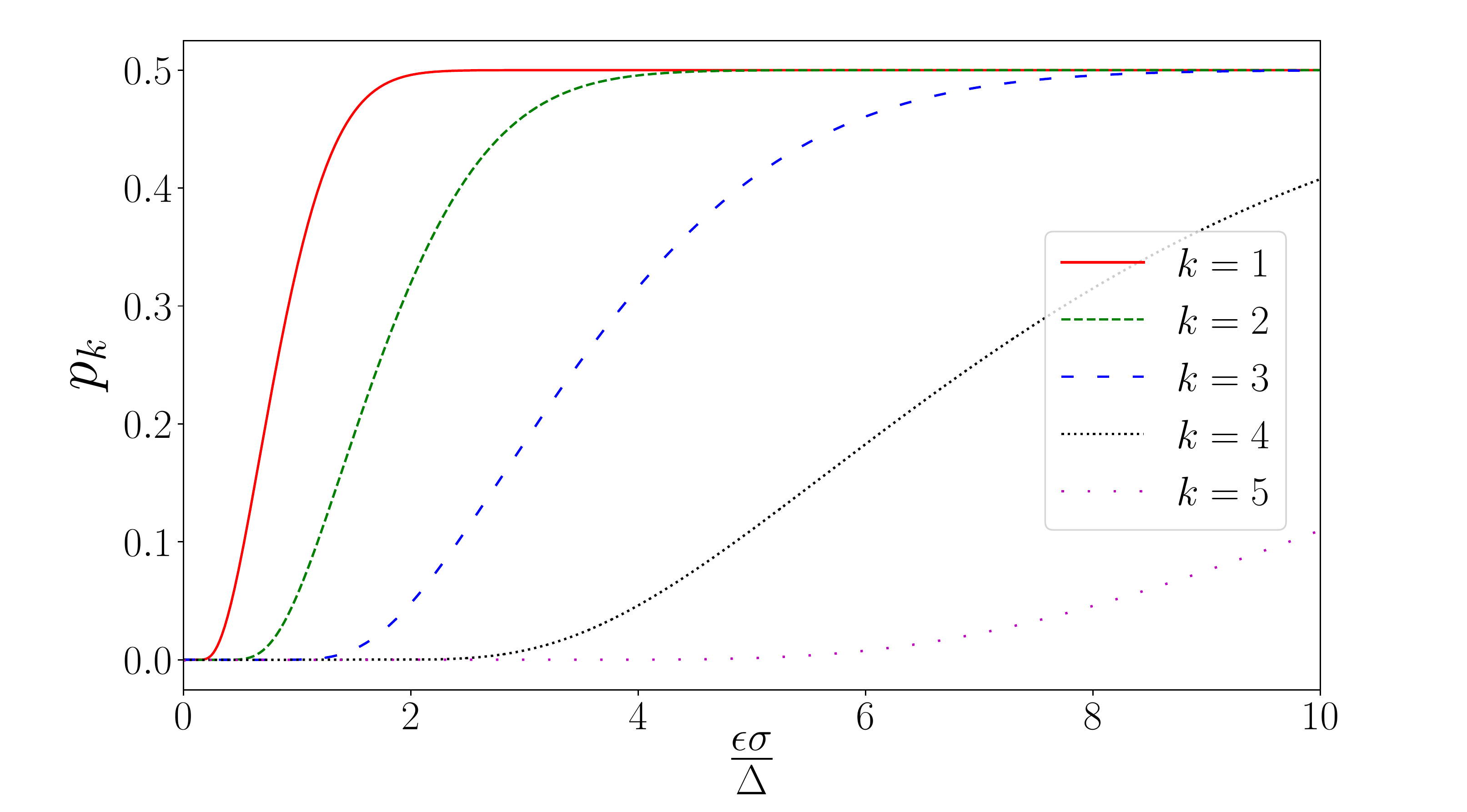}
	(c)\includegraphics[width=.3\linewidth]{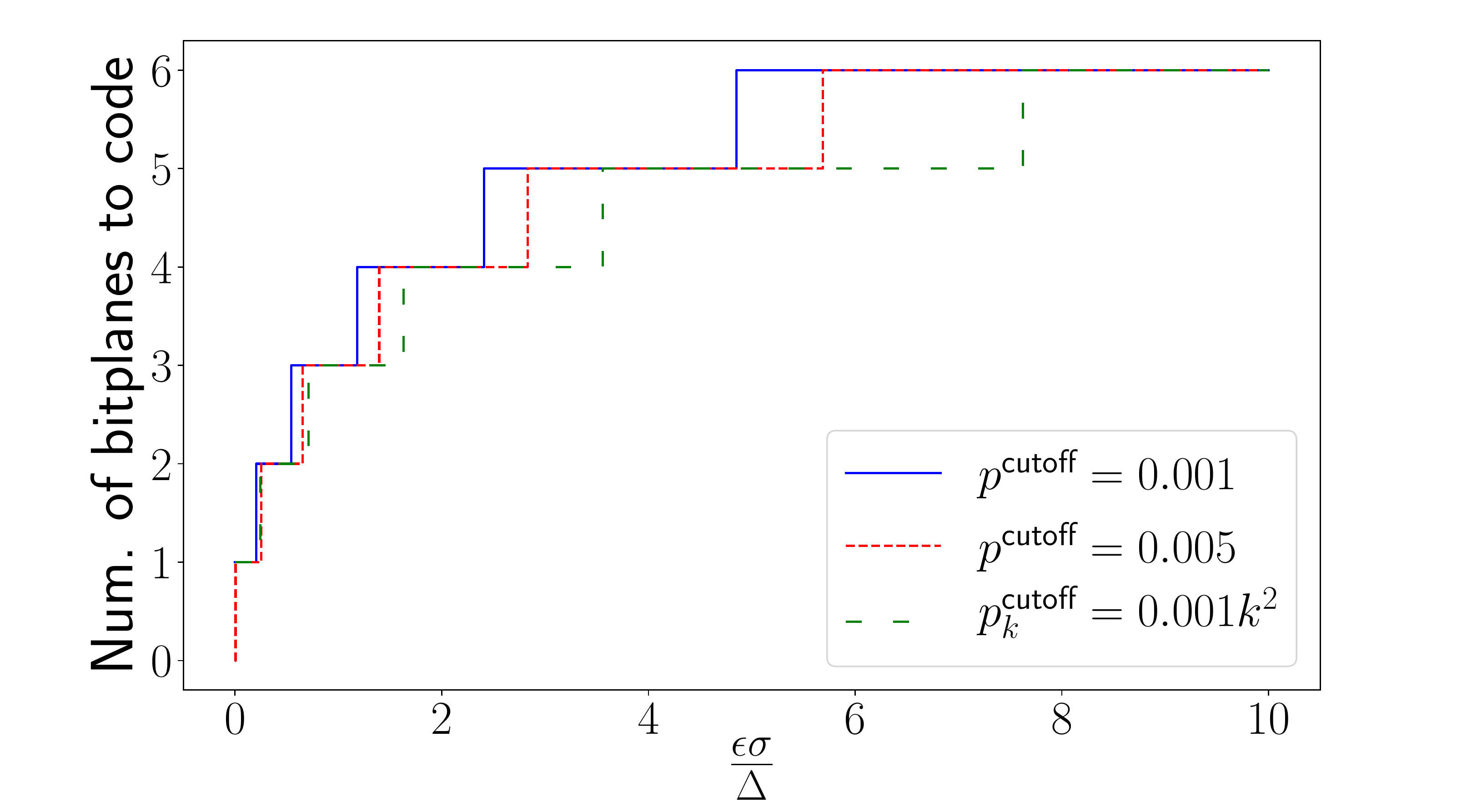}
	\caption{(a) The bit error likelihood $ L_{k} $ as a function of the distance parameter $ c $ when $ k = 3 $. (b) Bit error probability $ p_k $ as a function of the normalized prediction error $ \frac{\epsilon\sigma}{\Delta} $ for different values of $ k $. (c) The number of bitplanes that need to be coded as syndromes as a function of the normalized prediction error $ \frac{\epsilon\sigma}{\Delta} $.}
	\label{fig:reliability_behavior}
\end{figure*}

As expected, in our simulations we observe that this approach improves the LDPC
decoding process. In particular, it allows successful belief propagation using
higher code rates, compared to the agnostic prior, which translates to smaller
syndrome sizes and hence increased compression rate.

\subsection{Cutoff Probabilities for Syndrome Generation}\label{sec:cutoff_pr}
The behavior of the bit error probability, calculated according to
\eqref{eq:bit_error}, as a function of the normalized prediction error $
	\frac{\epsilon\sigma}{\Delta} $ for different values of $ k $ is shown in
Fig.~\ref{fig:reliability_behavior}(b). As the figure illustrates, the error
probability decreases sharply for bitplanes of higher significance. If the
error $\epsilon$ is sufficiently small, bitplanes beyond a certain significance
can be predicted error-free at the decoder, without the need to transmit
syndromes. In practice, it is reasonable to set a cutoff probability $
	p^{\text{cutoff}} $ below which bitplanes are not encoded.

This cutoff can be chosen in a number of ways. For example, we may choose a
fixed probability, e.g., $ p^{\text{cutoff}} = 0.001 $, or one that is
reciprocal to the syndrome length, i.e., $ p^{\text{cutoff}}\propto 1/m $,
where the constant of proportionality is chosen to ``control" the expected
frequency of bit errors in non-encoded bitplanes. Alternatively, since
bitplanes are decoded sequentially from LSB to MSB, we may consider assigning
different cutoffs to each bitplane, to prevent error propagation across
bitplanes. Cutoffs can therefore be assigned in a progressive manner from lower
to higher values, with lower significance bits treated more conservatively than
higher significance bits, e.g., $ p_{k}^{\text{cutoff}} = 0.001k $ or $
	p_{k}^{\text{cutoff}} \propto k/m $.

Fig.~\ref{fig:reliability_behavior}(c) shows the number of bitplanes that need
to be encoded as syndromes as a function of the normalized prediction error for
three different cutoff probabilities. The $ \Delta $ parameter controls the
quantizer resolution. Increasing $ \Delta $ is equivalent to using a coarser
scalar quantizer, which results in higher distortion, but makes bitplanes more
predictable. Thus, as the figure shows, fewer bitplanes need to be encoded as $
	\Delta $ increases, improving the compression rate. Hence, the $ \Delta $
parameter, which is one of the algorithm's design parameters, plays a key role
in controlling the rate-distortion trade-off.

For some bitplanes of low significance, the error rate $p_k$ is so high, that
the syndrome has effectively the same bit-size as the bitplane, i.e., requires
a rate 0 code. In those cases it makes more sense to simply transmit the
bitplane instead of spending the extra computation to code and decode it
without reducing the number of bits. Thus another cutoff can be set in practice
for the value of $p_k$, above which the bitplane is transmitted as is. This
cutoff is not as critical. Setting it lower than necessary makes the
compression slightly less efficient, but does not introduce any errors in the
iterative process.

For simplicity, we chose a fixed probability for both cutoffs. In our
experiments, typically 1 or 2 least significant bitplanes were transmitted as
is. At most 3 additional bitplanes were transmitted by syndrome coding at code
rates greater than 0 but less than 1. Syndrome coding and transmission of the
remaining, higher significance, bitplanes was unnecessary, as they could be reliably recovered at the
decoder from the source prediction.

\subsection{Encoding Complexity}
In general, the computational and storage cost of the linear measurement
operation is $\mathcal{O}(nm)$, dominated by the matrix-vector multiplication
$\mA\vx$. However, using a fast-transform-based measurement operator, as
described above, the computational cost can reduce to $\mathcal{O}(n\log n)$
and storage to $\mathcal{O}(n)$. The measurements are quantized using an $
	\mathcal{O}(n)$ scalar quantizer.

The generation of syndromes involves multiplication of the parity-check matrices
$ \mH_{(k)} $ of size $ m(1  - R_{k}) \times m $, where $ R_{k} \in \left[ 0,1
		\right] $ is the code rate used to encode the $ k^\mathrm{th} $ bitplane, by the
vectors of quantized measurements $ \vq_{(k)} $ of size $ m $. The worst case
complexity, when $ R \approx 0 $, is order $ \mathcal{O}(m^{2})$. However, in
practice, structured parity-check matrices can be used to generate syndromes
more efficiently, with complexity as low as $\mathcal{O}(m)$. Obtaining
the required code rates $ R_{k} $ involves calculating the bit error
probabilities $ p_{k} $ according to Thm.~\ref{thm:flip_prob}, which can be done
in constant time.

In summary, complexity is dominated by an $\mathcal{O}(n\log n)$ matrix-vector multiplication. Furthermore, the end-to-end encoder architecture
is very simple, as typical in distributed coding schemes, and can be
efficiently implemented in resource-constrained environments. Instead the
complexity has been transferred to the decoder, which has the ability to
exploit signal models through the prediction and the regularization process,
even if these signal models are too expensive to be computed at the encoder.
\section{Application: Multispectral Image Compression}
\label{sec:results}
As an example use of our approach, we demonstrate its application to
lightweight multispectral image compression targeted to on-board satellite
systems.

Multispectal images comprise of four to eight spectral bands, which are
typically highly correlated. Such images are often acquired on board satellites
for remote sensing applications such as mineral exploration, surveillance and
cartography. Advances in modern sensing technology have resulted in increases
in the resolution and quality of the sensing instruments, with a corresponding
increase in the data size. Thus, considering the communication constraints in
space, compression has become a necessity for such systems. Furthermore,
resource constraints make existing transform coding-based approaches, such as
JPEG-2000, unsuitable. Hence, there is significant interest in rate-efficient
compression algorithms with low-complexity encoders. For example, some
low-complexity multispectral compression methods can be found in~\cite{Rane,
	Wang, Abrardo, RateControl, Valsesia,VB_ITW16}.

The approach we present here expands on the approach in~\cite{Maxim}, by
incorporating likelihood estimation in syndrome decoding and further exploiting
inter-band correlations using successive prediction from already decoded bands.
Consistent with~\cite{Rane, Wang, Valsesia,VB_ITW16}, we consider images
comprising of four spectral bands $\vx_0,\ldots,\vx_3$, corresponding to blue,
green, red and infrared, respectively. We assume that the blue band, $\vx_0$,
is fully available at the decoder as side information and we explore the
compression of the remaining three spectral bands. The approach we present can
be naturally generalized to the compression of multiple correlated sources in
other modalities.

\subsection{Encoding}
\label{subsec:sim_encoding}
\begin{figure*}[t]
	\centering
	\includegraphics[width=\linewidth]{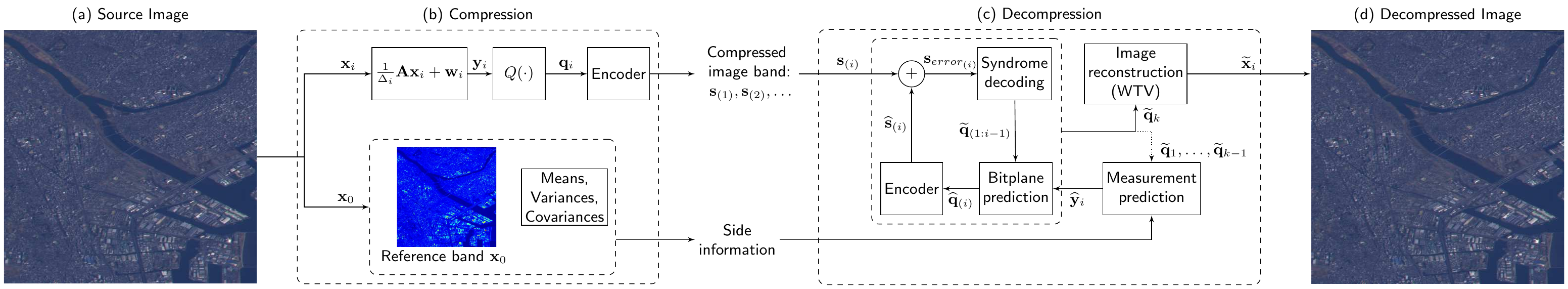}
	\caption{End-to-end system for the compression of a multispectral image. The source image (a) is compressed by the encoder (b) to produce syndromes and side information. The decoder (c) decodes the syndromes using the side information to produce the decompressed image (d).}
	\label{FIG:MSI_compression}
\end{figure*}

To encode the images, we first separate them to non-overlapping blocks of size
$n = 64\times 64$, which are treated and compressed independently using the
system in Fig.~\ref{FIG:MSI_compression}(b). Each block is measured with $ m =
	4000 $ measurements for each image band, $ \vx_{i}, i = 1,2,3 $, according to
\eqref{eq:measurement}, using the SRHT and an appropriate choice of $\Delta$ to
achieve the desired rate. Preliminary experiments demonstrated that very low
undersampling, with $ m = 4000 $, provided better rate-distortion performance
and allowed for better rate-distortion control through the choice of $\Delta$.
This is consistent with some of the findings in~\cite{Laska}. The measurements
are quantized with a $ B = 11 $ bits scalar quantizer.

In addition to syndromes, the encoder calculates statistics---specifically, the
mean and variance of each band, and the covariance with the other
bands---which are communicated to the decoder as part of the side information.
These statistics allow the encoder to estimate the prediction error and
determine the appropriate rate point for the syndromes. The statistics also enable the
decoder to predict the source.

Due to the complexity of generating codes at arbitrary rates during operation,
we maintain and select codes from a database of LDPC codes corresponding to
rates $ 0.05, 0.1, \dots, 0.95 $ of block length $ m = 4000 $, stored on-board.
Due to the effects of finite block lengths, to ensure accurate decoding, we
heuristically back off and select a code with rate $ 0.05 $ lower than the one closest to
capacity. For example, if for some bitplane $ k $ the error probability is $
	p_{k} = 0.11 $, the resulting channel capacity is $ C = 0.5 $ and we select
code rate $ 0.45 $ to encode. Using this heuristic, we observed that belief propagation converged almost always in practice, with extremely rare failures. Without the back-off, we observed an increase in LDPC decoding errors, leading to deterioration in reconstruction quality.

More sophisticated approaches might improve code rate selection, and resulting
performance. For example, recent work on finite blocklength codes derives
improved bounds for rate selection~\cite{Polyanskiy10,CIT-086}. In our
experiments, the additional complexity provided limited benefit. However, the
benefit might be greater in applications with smaller block lengths.

\subsection{Decoding}
The decoder, shown in Fig.~\ref{FIG:MSI_compression}(c), uses the
syndromes and the side information, i.e., the reference blue band $\vx_0$ and
the signals statistics, to recover the remaining spectral bands.

\subsubsection{Image Prediction}
To predict measurements at the decoder, we explore two approaches of
different complexity and predictive power.

\paragraph{Linear Prediction}   
The simplest approach is to use a linear minimum mean-square error (LMMSE) estimator to predict each of the three
bands $ \vx_{i}, i = 1,2,3 $ from the blue band $ \vx_{0} $:
\begin{equation} \label{eq:LMMSE}
	\vxh_{i} = \frac{\overline{\sigma}_{\vx_{0}\vx_{i}}}{\overline{\sigma}_{\vx_{0}}^{2}}\left( \vx_{0} - \overline{\mu}_{\vx_{0}}\mathbbm{1} \right) + \overline{\mu}_{\vx_{i}}\mathbbm{1},
\end{equation}
where $ \mathbbm{1} $ is the all-ones vector. Accordingly, the encoder must
calculate and communicate to the decoder the mean of each band
and its covariance with the blue band:
\begin{align*}
	\overline{\mu}_{\vx_{i}}           & = \frac{1}{n}\sum_{j=1}^n(\vx_{i})_{j},                                                                                      
	\\
	\overline{\sigma}_{\vx_{0}\vx_{i}} & = \frac{1}{n}\sum_{j=1}^n\left((\vx_{0})_j-\overline{\mu}_{\vx_{0}}\right)\left((\vx_{i})_j-\overline{\mu}_{\vx_{i}}\right), 
\end{align*}
where $ (\vx_{i})_{j} $, $ j=1,\ldots,4096 $, represents the $ j^\mathrm{th} $ pixel value of the $ i^\mathrm{th}
$ image band. Note that $ \overline{\mu}_{\vx_{0}} $ and $
	\overline{\sigma}_{\vx_{0}}^{2} $ do not need to be transmitted, as they can be
directly calculated at the decoder from $ \vx_{0} $. The coding overhead due to
transmission of these parameters is small. For example, using 16 bits per
parameter, the 3 covariances and 3 means require $6\times 16=96$ bits per
$64
	\times 64 $ block, an overhead of 0.00781 bits per pixel (bpp).

The prediction error for each band, which is also available at the encoder, is
equal to
\begin{equation}
	\epsilon_{i}^{2} = \|\vx_{i} - \vxh_{i}\|_{2}^{2} = n\left(\overline{\sigma}_{\vx_{i}}^{2} - \frac{\overline{\sigma}_{\vx_{0}\vx_{i}}^{2}}{\overline{\sigma}_{\vx_{0}}^{2}}\right).
\end{equation}

\paragraph{Successive Prediction}
\label{subsubsec:suc_pred}
Since the decoder may recover the measurements of each band one at a time, it
has additional information that can be used to predict subsequent bands. For
example, the green band is spectrally closer to the red band than the blue band
is, so, presumably, a successfully decoded green band contains additional
information that can assist in the prediction of the red band. Thus, successive
prediction leverages the results of all the previously decoded bands to improve
the prediction error.

The source signals $\vx_i$ are only available approximately at the decoder,
with an unknown error that is difficult to quantify due to the non-linearity of
the reconstruction process. On the other hand, their quantized measurements $
	\vq_{i}=Q(\vy_{i})$ are available exactly, i.e., assuming, as before, that
measurement reconstruction is reliable with $ \vqt_{i}=\vq_{i}$, and the
quantization error can be exactly quantified thanks to the effects of dither.
Thus, the already recovered measurements can be used for prediction and error
estimation.

Measurements of band $ i $, acquired according to \eqref{eq:measurement}, are
\begin{equation}
	\vy_{i} = \frac{1}{\Delta_{i}}\mA\vx_{i} + \vw_{i}.
\end{equation}
Furthermore, we use $\vyt_{i} = \vy_{i} - \vw_{i}$ to denote the pre-dithered
measurements, and $\vyt_{i}^{q} = Q(\vy_{i}) - \vw_{i}$ to denote the quantized
measurements with dither removed after quantization. Since dither coefficients
are drawn uniformly in $[-1,0)$, the quantization error $\ve_{i} = Q(\vy_{i}) -
	\vy_{i} = \vyt_{i}^{q} - \vyt_{i}$ also comprises of i.i.d. coefficients,
uniformly distributed in $[-\frac{1}{2},\frac{1}{2})$ \cite{Schuchman}. Thus,
the quantization error is a random variable $E_i$ with mean $ \mu_{E_{i}} = 0 $
and variance $ \sigma_{E_{i}}^{2} = \frac{1}{12} $.

Successful decoding of all bitplanes of band $ i $, provides the decoder with
the quantized measurements $\vq_{i}$ and, consequently, $ \vyt_{i}^{q} $.
Instead of linearly predicting the measurements from the blue band only,
successive prediction uses the quantized and dither-removed measurements of
bands $i=1,\ldots,k-1$, i.e., $ \vyt_{i}^{q} $, to predict the pre-dithered
measurements of band $k$ using a multi-variable LMMSE estimator:
\begin{align} \label{eq:succ_pred}
	\left(\vyth_{k}\right)_{j} & = C_{k,k-1}^{T}C_{k-1}^{-1}
	\begin{bmatrix}\left(\vyt_{0}\right)_{j} - \overline{\mu}_{\vyt_{0}}         \\
		\left(\vyt_{1}^{q}\right)_{j} - \overline{\mu}_{\vyt_{1}^{q}} \\
		\vdots                                                        \\
		\left(\vyt_{k-1}^{q}\right)_{j} - \overline{\mu}_{\vyt_{k-1}^{q}}\end{bmatrix} + \overline{\mu}_{\vyt_{k}},
\end{align}
for $ j = 1,\dots,m $ and where
\begin{equation}
	C_{k,k-1} = \begin{bmatrix}
		\overline{\sigma}_{\vyt_{k}\vyt_{0}} ,
		\overline{\sigma}_{\vyt_{k}\vyt_{1}^{q}} ,
		\dots,
		\overline{\sigma}_{\vyt_{k}\vyt_{k-1}^{q}}
	\end{bmatrix}^T,
\end{equation}
and
\begin{equation}
	C_{k-1} = \begin{bmatrix}
		\overline{\sigma}_{\vyt_{0}}^{2}           & \overline{\sigma}_{\vyt_{0}\vyt_{1}^{q}}       & \dots  & \overline{\sigma}_{\vyt_{0}\vyt_{k-1}^{q}}     \\
		\overline{\sigma}_{\vyt_{1}^{q}\vyt_{0}}   & \overline{\sigma}_{\vyt_{1}^{q}}^{2}           & \dots  & \overline{\sigma}_{\vyt_{1}^{q}\vyt_{k-1}^{q}} \\
		\vdots                                     & \vdots                                         & \ddots & \vdots                                         \\
		\overline{\sigma}_{\vyt_{k-1}^{q}\vyt_{0}} & \overline{\sigma}_{\vyt_{k-1}^{q}\vyt_{1}^{q}} & \dots  & \overline{\sigma}_{\vyt_{k-1}^{q}}^{2}
	\end{bmatrix}.
\end{equation}
To predict $ \vy_{k} $ the decoder therefore requires $ \vy_{0} $ (which is
available as side information) and $ \vq_{1}, \dots, \vq_{k-1} $ (which are
recovered by the decoder prior to predicting $ \vy_{k} $).

Note that \eqref{eq:succ_pred} implies that statistical parameters are
available for non-quantized measurements (e.g., $ \overline{\mu}_{\vyt_{k}} $), quantized
measurements (e.g., $ \overline{\mu}_{\vyt_{k-1}^{q}} $) and a mix of both (e.g., $
	\overline{\sigma}_{\vyt_{0}\vyt_{1}^{q}} $), which might not always be available at the
encoder. However, the dither allows us to compute the statistics of quantized
measurements from the non-quantized ones. If $ \wtY_{i} $ and $ \wtY_{i}^{q} $
denote the random variables corresponding to the entries of $ \vyt_{i} $ and $
	\vyt_{i}^{q}$, respectively, then
\begin{equation}
	\mu_{\wtY_{i}^{q}} = \E{[\wtY_{i}^{q}]} = \E{[\wtY_{i} + E_{i}]} = \E{[\wtY_{i}]} = \mu_{\wtY_{i}},
\end{equation}
since $E_i$ is zero mean, and
\begin{align}
	\sigma_{\wtY_{j}^{q}}^{2} & = \Var{[\wtY_{j}]} + \Var{[E_{j}]} = \sigma_{\wtY_{j}}^{2} + \frac{1}{12},
\end{align}
since $ E_{j} $ and $ \wtY_{j} $ are independent. Furthermore,
\begin{align}
	\sigma_{\wtY_{i}^{q}\wtY_{j}} & = \E{[(\wtY_{i} + E_{i})\wtY_{j}]} - \E{[\wtY_{i}^{q}]}\E{[\wtY_{j}]} = \sigma_{\wtY_{i}\wtY_{j}},
\end{align}
and, similarly
\begin{align}
	\sigma_{\wtY_{i}^{q}\wtY_{j}^{q}} & = \E{[(\wtY_{i} + E_{i})(\wtY_{j} + E_{j})]} - \E{[\wtY_{i}^{q}]}\E{[\wtY_{j}]} = \sigma_{\wtY_{i}\wtY_{j}}.
\end{align}
Thus, empirical statistics of quantized measurements can be approximated using
statistics of non-quantized measurements:
\begin{align}
	\overline{\mu}_{\vyt^{q}} \approx \overline{\mu}_{\vyt}, ~~                               & ~~\overline{\sigma}_{\vyt^{q}}^{2} \approx \overline{\sigma}_{\vyt}^{2} + \frac{1}{12},\label{eq:mean_appox}        \\
	\overline{\sigma}_{\vyt_{i}^{q}\vyt_{j}} \approx \overline{\sigma}_{\vyt_{i}\vyt_{j}}, ~~ & ~~ \overline{\sigma}_{\vyt_{i}^{q}\vyt_{j}^{q}} \approx \overline{\sigma}_{\vyt_{i}\vyt_{j}} \label{eq:cov_approx}.
\end{align}

Assuming approximations~\eqref{eq:mean_appox}--\eqref{eq:cov_approx} are used,
the encoder must calculate and communicate to the decoder the set $ \lbrace
	\overline{\mu}_{\vyt_{i}}, \overline{\sigma}_{\vyt_{i}}^{2},
	\overline{\sigma}_{\vyt_{i}\vyt_{j}} : i,j = 0,1,2,3, i \neq j \rbrace$, except
for $\overline{\sigma}_{\vyt_{3}}^{2}$, which is not needed, and $
	\overline{\mu}_{\vyt_{0}} $ and $ \overline{\sigma}_{\vyt_{0}}^{2} $, which are
used but not transmitted because, as before, they can be computed at the
decoder. Thus, 11 parameters need to be transmitted, with transmission
overhead, assuming again 16 bits per coefficient, equal to $11\times 16=176$
bits, i.e., 0.01432 bpp.

\begin{table*}[h]
	\caption{Decoding PSNR at 2 bpp ($ 512 \times 512 $ image crop)}
	\resizebox{\textwidth}{!}{
		\begin{tabular}[t]{@{}lccccccccc@{}}\toprule
			                                                                                          & \multicolumn{3}{c}{\textbf{PSNR (dB)}} & \phantom{abc} & \multicolumn{5}{c}{\textbf{BPP}}                                                                                               \\
			\cmidrule{2-4} \cmidrule{6-10}
			\textbf{}                                                                                 & \textbf{green}                         & \textbf{red}  & \textbf{infrared}                &  & \textbf{green} & \textbf{red} & \textbf{infrared} & \textbf{overhead} & \textbf{overall} \\ \toprule
			\multicolumn{1}{c}{\textbf{Benchmark}~\cite{Valsesia}}                                    & 37.79                                  & 32.76         & 34.24                            &  & 2.00           & 2.00         & 2.00              & ---               & 2.00             \\ \midrule
			\multicolumn{1}{c}{\textbf{Linear prediction}}                                                                                                                                                                                                                                      \\
			Prediction only                                                                           & 33.46                                  & 28.53         & 27.52                            &  & ---            & ---          & ---               & ---               & ---              \\
			$\Delta_\mathrm{green}=\Delta_\mathrm{red}=\Delta_\mathrm{infrared}=10.395$               & 39.52                                  & 38.89         & 39.52                            &  & 1.51           & 2.19         & 2.28              & 0.00781
			                                                                                          & 2.00                                                                                                                                                                                    \\
			$\Delta_\mathrm{green}=6.995; \Delta_\mathrm{red}=12.275;\Delta_\mathrm{infrared}=13$     & 42.19                                  & 38.10         & 38.15                            &  & 1.99           & 1.99         & 2.00              & 0.00781           & 2.00             \\ \midrule
			\multicolumn{1}{c}{\textbf{Successive prediction}}                                                                                                                                                                                                                                  \\
			$\Delta_\mathrm{green}=\Delta_{red}=\Delta_\mathrm{infrared}=9.925$                       & 39.79                                  & 39.09         & 39.66                            &  & 1.55           & 2.22         & 2.17              & 0.01432           & 2.00             \\
			$\Delta_\mathrm{green}=7.055; \Delta_\mathrm{red}=12.05; \Delta_\mathrm{infrared}=11.675$ & 42.12                                  & 38.21         & 38.92                            &  & 1.98           & 1.99         & 1.98              & 0.01432           & 2.00             \\
			\bottomrule
		\end{tabular}}
	\label{table:results1}
\end{table*}

The encoder can calculate the prediction mean squared error (MSE) to estimate
the rates required for the syndromes. The prediction MSE is
\begin{align}
	\mathrm{MSE}_{k} & = \frac{1}{m}\|\vyth_{k} - \vyt_{k}\|_2^2= \overline{\sigma}_{\vyt_{k}}^{2} - C_{k,k-1}^{T}C_{k-1}^{-1}C_{k,k-1}.
\end{align}
The dithered prediction is obtained by adding the dither to the pre-dithered
prediction i.e., $ \vyh_{k} = \vyth_{k} + \vw_{k} $. Thus, the prediction error
is
\begin{equation}
	\epsilon_{\vy_{k}}^{2} = \|\vy_{k} - \vyh_{k} \|_2^{2} = \|\vyth_{k} - \vyt_{k}\|_2^2.
\end{equation}
To calculate the code rates for the syndromes as per Thm.\ref{thm:flip_prob},
we need to determine the distance $ \epsilon_{\vx_{k}} = \|\vx_{k} - \vxh_{k}
	\|_2 $ between a source $ \vx_{k} $ and its prediction $ \vxh_{k} $. We use the
RIP \cite{CandesTao1} to approximately relate the distance of the measurements
and their predictions, $ \epsilon_{\vy_{k}} $, to $ \epsilon_{\vx_{k}} $ as
follows
\begin{equation}
	\epsilon_{\vx_{k}}^{2} \approx \frac{n\Delta^{2}}{m}\epsilon_{\vy_{k}}^{2}~\label{eq:RIP_MSE_estimate}.
\end{equation}
Given that $\mA$ is almost square and unitary, the RIP constants are very
small, so the estimate in~\eqref{eq:RIP_MSE_estimate} is accurate.

\begin{table*}[h]
	\caption{Decoding PSNR at 1.68 bpp (full $ 7040 \times 7936$ image)}
	\resizebox{\textwidth}{!}{
		\begin{tabular}[t]{@{}lccccccccc@{}}\toprule
			                                                                                        & \multicolumn{3}{c}{\textbf{PSNR (dB)}} & \phantom{abc} & \multicolumn{5}{c}{\textbf{BPP}}                                                                                                   \\
			\cmidrule{2-4} \cmidrule{6-10}
			\textbf{}                                                                               & \textbf{green}                         & \textbf{red}  & \textbf{infrared}                &      & \textbf{green} & \textbf{red} & \textbf{infrared} & \textbf{overhead} & \textbf{overall} \\ \toprule
			\multicolumn{1}{c}{\textbf{Benchmark}~\cite{Valsesia}}                                  & 39.06                                  & 37.60         & 35.80                            &      & 1.68           & 1.68         & 1.68              & ---               & 1.68             \\ \midrule
			\multicolumn{1}{c}{\textbf{Linear prediction}}                                                                                                                                                                                                                                        \\
			Prediction only                                                                         & 37.05                                  & 31.67         & 27.32                            &      & ---            & ---          & ---               & ---               & ---              \\
			$\Delta_\mathrm{green}=\Delta_\mathrm{red}=\Delta_\mathrm{infrared}=9.095$              & 41.84                                  & 41.07         & 40.08
			                                                                                        &                                        & 1.17          & 1.70                             & 2.15 & 0.00781
			                                                                                        & 1.68                                                                                                                                                                                        \\
			$\Delta_\mathrm{green}=5.225; \Delta_\mathrm{red}=9.4;\Delta_\mathrm{infrared}=14.385$  & 44.96                                  & 40.89         & 37.85                            &      & 1.67           & 1.67         & 1.68              & 0.00781           & 1.68             \\ \midrule
			\multicolumn{1}{c}{\textbf{Successive prediction}}                                                                                                                                                                                                                                    \\
			$\Delta_\mathrm{green}=\Delta_\mathrm{red}=\Delta_\mathrm{infrared}=8.7325$             & 42.08                                  & 41.27         & 40.28                            &      & 1.20           & 1.70         & 2.09              & 0.01432           & 1.68             \\
			$\Delta_\mathrm{green}=5.275; \Delta_\mathrm{red}=8.95; \Delta_\mathrm{infrared}=13.15$ & 44.91                                  & 41.13         & 38.32                            &      & 1.66           & 1.66         & 1.66              & 0.01432           & 1.68             \\
			\bottomrule
		\end{tabular}}
	\label{table:results2}
\end{table*}

\subsubsection{Image Recovery}
Once all bitplanes have been successfully decoded, the quantized measurements
$\vqt_{i}, i = 1,2,3, $ are used to reconstruct the image by solving the sparse
optimization problem \eqref{eq:opt_general} with an appropriately chosen data
fidelity penalty and regularizer. In our experiments we found that, for our
particular rate points of interest with low-distortion, the quadratic penalty
performed better than penalties promoting consistent reconstruction. This,
however might not be the case if the chosen operating point requires coarser
quantization.

In reconstruction, similar to~\cite{weightedl1}, we exploit the presence of the
blue reference to weight the regularizer. In particular, we use a total
variation (TV)-based regularizer, known to enforce a good model for images by
imposing sparsity in their gradient. However, we also expect that if an edge is
present in the blue reference band, the edge is likely to be present in the
reconstructed band as well, i.e., at the same pixel locations there is going to
be a large gradient spike that should not be penalized. To incorporate this
additional information on the gradient from the blue reference, we use
weighted-TV (WTV),
\begin{multline}
	\mathcal{R}_\mathrm{WTV}(X) = \\\sum_{s,t} \sqrt{W_{s,t}^{x}(X_{s,t} - X_{s-1,t})^{2} + W_{s,t}^{y}(X_{s,t} - X_{s,t-1})^{2}},
	\label{eq:wtv}
\end{multline}
where $X$ is a 2D image, $W^x$ and $W^y$ are 2D sets of weights, and $(s,t)$
are image coordinates. Larger weights penalize edges more in the respective
location and direction, while smaller weights reduce their significance. When
all the weights are set to 1, the regularizer is the standard TV regularizer.

The regularizer weights at pixel spatial coordinate $(s,t) $ are chosen based
on the norm of the gradient of the blue reference band at the corresponding
pixel, denoted $\Phi(X_{0_{s, t}})$, which is computed using
\begin{equation}
	\Phi(X_{0_{s, t}}) = \sqrt{(X_{0_{s, t}} - X_{0_{s-1, t}})^{2} + (X_{0_{s, t}} - X_{0_{s, t-1}})^{2}},
\end{equation}
where $ \refX s t $ is the value of the image pixel of the reference band $
	\vx_{0} $ at coordinate $(s,t) $.

The weights are then computed according to
\begin{equation}
	W_{s,t}^{x} = W_{s,t}^{y}=\left\{
	\begin{array}{rl}
		0.2, & \mathrm{if}~\Phi(\refX s t)> \tau, \\
		1,   & \mathrm{otherwise},
	\end{array}
	\right.,
\end{equation}
where $ \tau = 0.3 $ is an experimentally tuned threshold, qualifying which
gradient norms are considered to be significant and likely to be present in
other bands.

To summarize, the decoder solves
\begin{align}
	\vxt_{i} = \arg\min_{\vx}\bigl\|\vqt_{i}-\frac{1}{\Delta_{i}}\mA\vx-\vw_{i}\bigr\|_2^2+\lambda
	\mathcal{R}_\mathrm{WTV}(\vx),
	\label{eq:weighted_wtv_ell2}
\end{align}
where $\lambda=0.1$ was tuned experimentally using a small part of the data.
Several ways exist to solve~\eqref{eq:weighted_wtv_ell2}; we use a fast
iterative shrinkage thresholding (FISTA)-based
approach~\cite{kamilov2017parallel}.

\subsection{Simulation Results}
To measure the performance of our method we consider the trade-off between the
compression rate and the quality of reconstruction, measured using the peak
signal-to-noise ratio (PSNR) metric. In decibels (dB), PSNR is defined as
\begin{equation}
	\text{PSNR}(\vx_{i}, \vxt_{i}) = 10 \log_{10} \left( \frac{\text{max}(\vx_{i})^{2}}{\text{MSE}(\vx_{i}, \vxt_{i})}\right),
\end{equation}
where $ \text{MSE}(\vx_{i}, \vxt_{i}) $ is the mean squared error between the
source $ \vx_{i} $ and its reconstruction $ \vxt_{i} $ and $
	\text{max}(\vx_{i}) $ returns the value of the largest element of $ \vx_{i} $.
As a benchmark, we use the results in~\cite{Valsesia} where a similar
complexity encoder was used under the same settings to compress 4-band images
acquired by the ALOS satellite~\cite{jaxa}. We performed tests on an entire $
	7040 \times 7936$ image, as well as more extensive testing on a smaller $ 512
	\times 512 $ patch of the same image. This patch, shown in
Fig.~\ref{FIG:MSI_compression}(a), was chosen as it was deemed challenging to
compress.

The encoding parameters, $\Delta$ in particular, were chosen to match the rates
used in~\cite{Valsesia}. We considered image blocks of size $ n = 64 \times 64
$, and examined the performance under both the simple linear prediction and the
successive prediction approaches above. Note that our preliminary
work~\cite{Maxim} only considers the simple linear prediction approach and does
not exploit likelihood computation using Thm.~\ref{thm:likelihood} in the
decoding.

The results for the compression of the image patch in
Fig.~\ref{FIG:MSI_compression}(a) are reported in Table~\ref{table:results1}.
The first row, titled {\em Prediction only}, lists the quality of the linear
prediction using only side information and prediction parameters, i.e., without
syndrome decoding and reconstruction. This quantifies the quality of predicting
each band from the reference blue band and serves as a baseline. As expected,
the quality of prediction matches the spectral distances between the predicted
bands and blue band: the closest green band is easiest to predict, followed by
the red band and the furthest infrared band.

We experimented with two different approaches to set the quantization
resolution parameter $ \Delta $, which controls the rate-distortion trade-off
as discussed in Sec.~\ref{sec:cutoff_pr}. In the first approach we chose the
same $ \Delta $ for all 3 bands, such that the average compression rate is the
target of 2 bpp. As expected, the reconstruction quality is fairly similar
among the bands, while the bit budget is spread unevenly, with the easier to
predict bands requiring lower code rates. In the second approach, the $ \Delta
$ was set such that each band is compressed at the target rate of 2 bpp. In
this case, as expected, the reconstruction quality improves for the easier to
predict bands. This improvement comes at the expense of the harder to predict
bands, whose reconstruction quality deteriorates. Note that the effectively
negligible overhead associated with transmitting the statistical parameters at
16 bits per parameter is included in a separate ``overhead'' column, and not
counted in the calculation of the reported per-band bpp values.

As is apparent from Table~\ref{table:results1} the approach significantly
outperforms the benchmark in terms of quality of reconstruction. The ranges of
improvements in reconstruction PSNR are summarized in the first two columns of
Table~\ref{table:gains1}. The performance improvements of linear and successive
predictions are quite similar, with successive prediction doing slightly better
than linear prediction in most cases.

A similar set of experiments was conducted for the larger $ 7040 \times 7936$
image with a target rate of 1.68 bpp, considering both linear and successive
prediction and both fixed and variable $ \Delta $. The results are shown in
Table~\ref{table:results2}. As expected, the results exhibit a similar behavior
to the results of the smaller image patch. A common $\Delta$ for all bands
leads to similar reconstruction quality at different compression rates, whereas
varying $\Delta$ for each band, such that the rate is the same, leads to
variations in reconstruction quality. The performance of the linear and
successive prediction is quite similar. As with the smaller image patch, the
proposed approach outperforms~\cite{Valsesia}. The improvements in
reconstruction quality are summarized in the two right columns of
Table~\ref{table:gains1}.

For the large $ 7040 \times 7936$ image, we also examined the quality of
reconstructed measurements $ \vqt_{i} $ with respect to decoding errors.
Table~\ref{table:BER} lists the resulting average bit error rates of $ \vqt_{i}
$, compared to the true $ \vq_{i} $ for the scenarios we simulated. We observe
that the measurements are reliably reconstructed, with bit error rates on the
order of $10^{-4}$. This provides some empirical support that our heuristic to
accommodate the effects of finite block lengths by choosing slightly
conservative code rates, described in Sec.~\ref{subsec:sim_encoding}, works
well in practice.

\begin{table}[t]
	\caption{Quality improvement over benchmark approach} 
	\centering 
	\begin{tabular}{c c c c c c} 
		\toprule 
		                    & \multicolumn{2}{c}{$ 512 \times 512 $ image
		}                   &                                             & \multicolumn{2}{c}{$ 7040 \times 7936$ image}                                                 \\
		                    & \multicolumn{2}{c}{2.00
		bpp}                &                                             & \multicolumn{2}{c}{1.68 bpp}                                                                  \\
		\cmidrule{2-3} \cmidrule{5-6}
		\textbf{Prediction} & \textbf{same} $ \Delta $                    & \textbf{variable} $
		\Delta $            &                                             & \textbf{same} $ \Delta $                      & \textbf{variable} $
			\Delta $                                                                                                                                                          \\ 
		\midrule 
		\textbf{Linear}     & 1.7--6.1dB                                  & 3.9--5.3dB                                    &                     & 2.8--4.3dB & 2.0--5.9dB \\ 
		\textbf{Successive} & 2.0--6.3dB                                  & 4.3--5.5dB                                    &                     & 3.0--4.5dB & 2.5--5.9dB \\
		\bottomrule
	\end{tabular}
	\label{table:gains1}
\end{table}

\begin{table}[t]
	\caption{Average bit error rate of reconstructed measurements (units are $\times10^{-4}$, for a 7040 $ \times $ 7936 image)} 
	\centering 
	\begin{tabular}{c c c c c c c c} 
		\toprule 
		                    & \multicolumn{3}{c}{\textbf{same} $ \Delta $}
		                    &                                              & \multicolumn{3}{c}{\textbf{variable} $\Delta $}                                                                            \\
		\cmidrule{2-4} \cmidrule{6-8}
		\textbf{Prediction} & \textbf{green}                               & \textbf{red}                                    & \textbf{infrared} &  & \textbf{green} & \textbf{red} & \textbf{infrared} \\
		\midrule 
		\textbf{Linear}     & 1.01                                         & 1.40                                            & 1.77              &  & 1.71           & 1.53         & 1.77              \\
		\textbf{Successive} & 1.15                                         & 1.51                                            & 2.00              &  & 1.76           & 1.95         & 2.00              \\
		\bottomrule
	\end{tabular}
	\label{table:BER}
\end{table}

\section{Discussion}
\label{sec:conclusion}
The proposed approach exploits advances in sampling theory and signal
representations, mostly due to the development of compressed sensing and sparse
signal recovery. From this area we inherit several desirable properties,
including the universality and the simplicity of the encoder, as well as the
ability to incorporate complex signal models and signal dependencies during
reconstruction. Because of that, the proposed approach is also future proof: an
improvement in signal models can improve decoding performance, without
requiring any change in the encoder.

Our development treads at the intersection of quantization, sparse signal
processing, and information theory. While we provide a flexible framework for
distributed coding, exploiting a modern signal processing paradigm, it is not
evident that our approach is in any sense optimal. Optimality in an
information-theoretic, rate-distortion sense is, in fact, difficult to
quantify. Modern signal models, such as sparsity and structured sparsity, do
not fit well the probabilistic information-theoretic framework typically used
in deriving optimality bounds. While some progress has been made, e.g.,
see~\cite{Kipnis}, understanding the information-theoretic properties of such
sources is still an open problem and an active area of research. More recently
developed learning-based signal models~\cite{Bora17} pose even more theoretical
questions.

A significant practical advantage of our approach is that the transmission rate
can be estimated at the transmitter using only an estimate of the $\ell_2$
error at the decoder in Thm.~\ref{thm:flip_prob}. Thus, given a fixed
quantization interval $\Delta$, the encoder can easily compute the transmission
rate. Furthermore, typically the reconstruction error is expected to be
approximately proportional to $\Delta$. A more complex encoder may take this
general trend into account, potentially in a rate-distortion optimization
component. However, due to the reconstruction's non-linear nature, the exact
relationship between $\Delta$ and the reconstruction error is still not well
understood, complicating the development of such a component.

A proper, theoretically-motivated, approach for trading off the values of $m$
and $ \Delta $ to achieve a certain rate-distortion trade-off remains an open
question of practical importance. While we tuned $m$ empirically in our
experiments, a more principled approach is desirable. A rate-distortion
function that takes $m$ and $\Delta$ into account would be ideal, but still
elusive.

\bibliographystyle{IEEEtran}
\bibliography{bibliography}

\appendices
\section{Proof of Theorem 1}
\label{appx:thm1_prf}
\begin{IEEEproof}
	Consider a single measurement of the signal $ \vx \in \mathbb{R}^{n} $ and its
	prediction $ \vxh \in \mathbb{R}^{n} $,
	\begin{equation}
		y = \frac{1}{\Delta}\inner\va\vx + w,~\why = \frac{1}{\Delta}\inner\va\vxh + w,
	\end{equation}
	where $ \va \in \mathbb{R}^{n} $ is the measurement vector, $ \Delta \in
		\mathbb{R^{+}} $ is a scaling parameter, $ w $ is dither, and \va, $\Delta$ and
	$w$ are common in both measurements. Assume entries of $ \va $ are drawn from
	an i.i.d.,  $\mathcal{N}(0,\sigma^2)$ distribution, and  $ w $ is drawn
	uniformly in $ [-1,0) $.  Let
	\begin{equation}
		Q_{n}(y) = q_{(n)}q_{(n-1)}\dots q_{(1)} \label{eq:Q_notation}
	\end{equation}
	represent the values of the first $ n $ LSBs of the quantized measurement $
		Q(y) $ of $ y $.

	\begin{figure}[b]
		\centering
		\includegraphics[width=.85\linewidth]{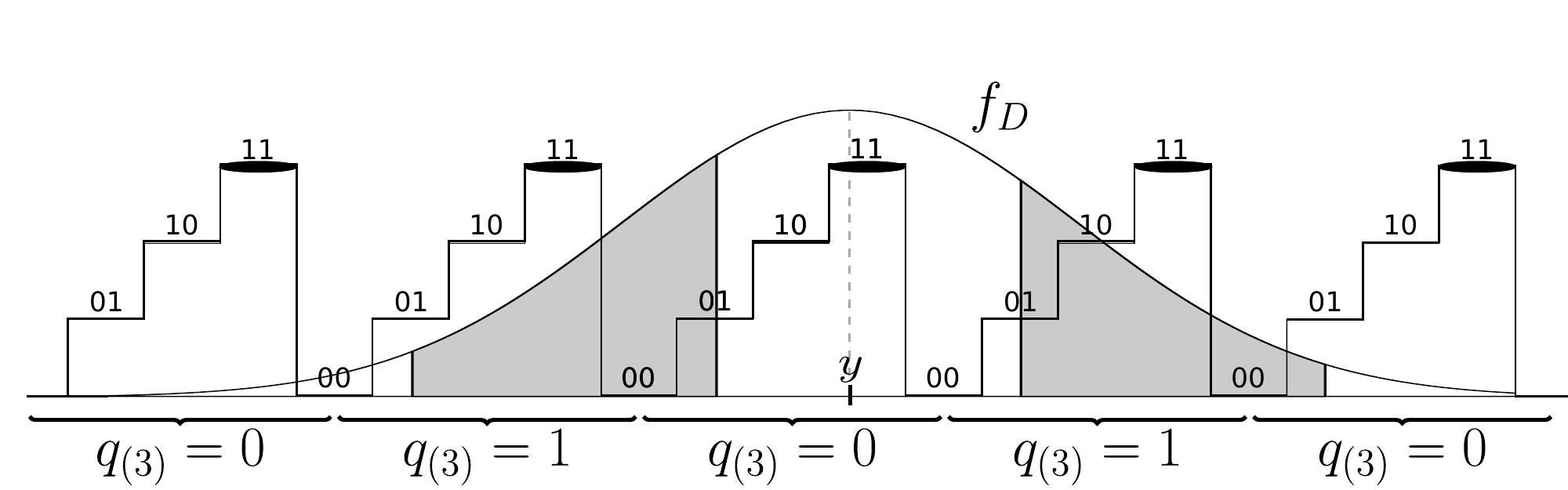}
		\caption{The error probability $ p_{3} $ corresponds to the area of the union of inconsistent intervals (shaded).}
		\label{fig:bit_flip_prob}
	\end{figure}

	The decoder predicts $ q_{(k)} $ based solely on $ \why $ and $ Q_{k-1}(y) $,
	which is assumed perfectly known or already correctly decoded. Knowledge of $
		Q_{k-1}(y) $ is equivalent to knowledge of $y$ quantized using a $(k-1)$-bit
	universal quantizer~\cite{universal}. Such a quantizer is equivalent to a
	uniform scalar quantizer with only the $k-1$ LSBs preserved, i.e., one with
	$2^{k-1} $ distinct levels, which repeat every $ 2^{k-1} $ quantization
	intervals. An example $2$-bit universal quantizer is shown in
	Fig.~\ref{fig:bit_flip_prob}. The value of the third bit, $ q_{(3)} $,
	alternates every $ 2^2 = 4 $ quantization intervals.

	Let $ \sI_{Q_{k-1}(y)} $ denote the set of all quantization intervals that map to $ Q_{k-1}(y) $, that is
	\begin{equation}
		\sI_{Q_{k-1}(y)} = \{\left[a,b\right) \mid Q_{k-1}(t) = Q_{k-1}(y)~\forall t \in \left[a,b\right) \}. \label{eq:I_notation}
	\end{equation}
	Similarly, $ \sI_{0Q_{k-1}(y)}~(\sI_{1Q_{k-1}(y)}) $ naturally extends the
	above definition for the case when the $ k^\mathrm{th} $ bit is $ 0~(1) $. In
	Fig.~\ref{fig:bit_flip_prob}, $ Q_{2}(y) = 11 $ therefore $
		\sI_{Q_{k-1}(y)}=\sI_{11}$ corresponds to the union of all the intervals shown
	in bold ellipses, while $ \sI_{0Q_{k-1}(y)}=\sI_{011}$ corresponds to the union
	of the intervals shown in bold ellipses only in the non-shaded regions.

	To generate a prediction $ \whq_{(k)} $ of $ q_{(k)} $, the decoder first finds
	the interval $ \sI^{\star} \in \sI_{Q_{k-1}(y)} $ with midpoint closest to $
		\why $:
	\begin{equation}
		\sI^{\star} = \underset{\sI \in \sI_{Q_{k-1}(y)}}{\mathrm{argmin}}\left| \why - \frac{\sup(\sI) + \inf(\sI)}{2} \right|.
	\end{equation}
The decoder then assigns $ \whq_{(k)} $ the value associated with the $ k $-bit quantization interval which contains $ \sI^{\star} $, that is
	\begin{equation}
		\whq_{(k)} =
		\begin{cases}
			0 & \text{if  } \exists \widetilde{\sI} \in \sI_{0Q_{k-1}(y)} \text{ such that  }\sI^{\star} \subset \widetilde{\sI} \\
			1 & \text{otherwise }
		\end{cases}.
	\end{equation}

	This bitplane prediction process is illustrated in
	Fig.~\ref{fig:bit_flip_prob}. If $ \why $ is inside the shaded regions, it will
	be closest to an interval $ \sI^{\star} \in \sI_{Q_{2}(y)} = \sI_{11} $ which
	is contained in an interval of $ \sI_{1Q_{2}(y)} = \sI_{111} $, thus predicting
	$ \whq_{(3)} = 1 $. Similarly, if $ \why $ is inside a non-shaded region, the
	decoder assigns $ \whq_{(3)} = 0 $.

	For a given $ y $, we call intervals of $ \Reals $ which would result in the $
		k^\mathrm{th} $ bit of $ \why $ to be mapped to $ \whq_{(k)} = q_{(k)} $,
	\textit{consistent} intervals. The width of any consistent interval is $
		2^{k-1} $ and the distance from its center to the center of a neighboring
	consistent interval is $ 2^{k} $. In Fig.~\ref{fig:bit_flip_prob}, consistent
	intervals correspond to the non-shaded regions, each of which has a width of $
		2^2=4 $. The union of all consistent intervals $ \sC(y) $ is given by
	\begin{align}
		\sC(y)
		 & = \lfloor y + \frac{1}{2} \rfloor + \bigcup\limits_{l \in \Integers}\left[-2^{k-2} + l2^{k}, l2^{k} + 2^{k-2} \right]. \label{eq:consistent_intervals}
	\end{align}
	The signed distance of the measurements, $ D $, is defined as
	\begin{align}
		D = y - \why = \frac{1}{\Delta}\inner\va{\vx-\vxh} = \frac{1}{\Delta}\sum_{i=1}^{n}\va_{i}\left(\vx_{i}-\vxh_{i}\right),
	\end{align}
	where $ \va_{i}, \vx_{i} $ and $ \vxh_{i} $ denote the $ i^\mathrm{th} $
	element of the $ \va, \vx $ and $ \vxh $ vectors, respectively. Since $D$ is a
	linear combination of $ n $ independent Gaussian random variables, it is also a
	random variable with Gaussian distribution with mean equal to
	\begin{align}
		\E{\left[D\right]} = \E{\left[\frac{1}{\Delta}\sum_{i=1}^{n}\va_{i}\left(\vx_{i}-\vxh_{i}\right)\right]} = \frac{1}{\Delta}\sum_{i=1}^{n}\E{\left[\va_{i}\right]}\!\left(\vx_{i}-\vxh_{i}\right) = 0,
	\end{align}
	where we recall that the $ \vx $ and $ \vxh $ are arbitrary, non-random,
	vectors. The variance of $ D $ equals
	\begin{align}
		\Var{\left(D\right)} & = \E{\left[D^{2}\right]} = \frac{1}{\Delta^{2}}\E{\left[\left(\sum_{i=1}^{n}\va_{i}\left(\vx_{i}-\vxh_{i}\right)\right)^{2}\right]}               \\
		                     & = \frac{1}{\Delta^{2}}\sum_{i=1}^{n}\E{\left[\va_{i}^{2}\right]}\left(\vx_{i}-\vxh_{i}\right)^{2} = \left(\frac{\sigma\epsilon}{\Delta}\right)^2.
	\end{align}
	In other words, $ D\sim
		\mathcal{N}\left(0,\left(\frac{\sigma\epsilon}{\Delta}\right)^2\right) $ and
	$f_{D}(\cdot)$ denotes its probability density function (pdf).

	For a given $ y $, the probability that $ \whq_{(k)} = q_{(k)} $ is the
	probability that $ \why \in \sC(y) $. This probability can be determined by
	integrating $ f_{D}(\cdot) $ over the coset $ \sC(y) - y $ as follows:
	\begin{equation}
		\Pr(\whq_{(k)} = q_{(k)}\mid y) = \int\limits_{u\in \sC(y)-y}f_{D}(u)du. \label{eq:f_D}
	\end{equation}
	In order to evaluate (\ref{eq:f_D}), we introduce the function $ g(\cdot) $,
	which is a rectangular function of width $ 2^{k-1} $, repeated at intervals of
	$ 2^{k} $, defined as
	\begin{equation}
		g(u) = \mathrm{rect}\left(\frac{u}{2^{k-1}}\right)\ast\sum_{l=-\infty}^{\infty}\delta\left(u-l2^{k}\right),
	\end{equation}
	where $ \delta(\cdot) $ is the Dirac delta function, $ \ast $ is the
	convolution operator on a pair of functions $ f $ and $ h $ defined as
	\begin{equation}
		\left(f \ast h\right)\left(u\right) = \int_{-\infty}^{\infty}f\left(\tau\right)h\left(u-\tau\right)d\tau
	\end{equation}
	and $ \mathrm{rect}\left(\cdot\right) $ is the rectangular function of unit
	width and height
	\begin{equation}
		\mathrm{rect}\left(u\right) =
		\begin{cases}
			0 & \text{for  } \left|u\right| > \frac{1}{2}   \\
			1 & \text{for  } \left|u\right| \le \frac{1}{2}
		\end{cases}.
	\end{equation}
	Noting that the support of $ g(u - \lfloor y + \frac{1}{2} \rfloor) $ is
	exactly equal to $ \sC(y) $, we evaluate \eqref{eq:f_D} as
	\begin{equation}
		\Pr(\whq_{(k)} = q_{(k)}\mid y) = \int_{-\infty}^{\infty}f_{D}\left(u\right)g\left(u-t\right)du, \label{eq:f_{D}_given_y}
	\end{equation}
	where $ t = \lfloor y + \frac{1}{2} \rfloor - y $. Note that the value of $ t $
	(rather than direct knowledge of $ y $) suffices for evaluation of the
	right-hand side of \eqref{eq:f_{D}_given_y}, so we can equivalently write $
		\Pr(\whq_{(k)} = q_{(k)}\mid y) = \Pr(\whq_{(k)} = q_{(k)}\mid t) $ i.e.,
	\begin{equation}
		\Pr(\whq_{(k)} = q_{(k)}\mid t) = \int_{-\infty}^{\infty}f_{D}\left(u\right)g\left(u-t\right)du, \label{eq:f_{D}_given_t}
	\end{equation}

	As $ y $ is a random variable and not known at the decoder, $ t $ is also a
	random variable. Due to the additive dither $ w $, the position of $ y $ is
	uniformly distributed across the quantization interval and therefore $ t $ is
	distributed uniformly in $ \left[-\frac{1}{2}, \frac{1}{2} \right]
	$\cite{Schuchman}. Thus, the pdf of $t$ is $ f_{t}\left(\cdot\right) =
		\mathrm{rect}(\cdot) $.

	To determine the probability that $ \whq_{(k)} = q_{(k)} $, we marginalize out the dependence in (\ref{eq:f_{D}_given_t}) on $ t $ i.e.,
	\begin{align}
		\Pr & \left(\whq_{(k)} = q_{(k)}\right) = \int_{-\frac{1}{2}}^{\frac{1}{2}}\Pr\left(\whq_{(k)} = q_{(k)}\mid \tau\right)f_{t}\left(\tau\right)d\tau                                                                                          \\
		    & = \int_{-\infty}^{\infty}\left[\int_{-\infty}^{\infty}f_{D}\left(u\right)g\left(u-\tau\right)du\right]f_{t}\left(\tau\right)d\tau \label{eq:t_supp}                                                                                    \\
		    & = \int_{-\infty}^{\infty}f_{D}\left(u\right)\left[\left(g \ast f_{t}\right)(u)\right]du                                                                                                                                                \\
		    & = \int_{-\infty}^{\infty}\mathcal{F}\left\lbrace f_{D}\left(u\right)\right\rbrace \left[\mathcal{F}\left\lbrace g\left(u\right)\right\rbrace\mathcal{F}\left\lbrace f_{t}\left(u\right)\right\rbrace\right]d\xi \label{eq:Plancherel},
	\end{align}
	where in \eqref{eq:t_supp} we note that the support of $ f_{t}(\cdot) $ is only
	$ \left[-\frac{1}{2}, \frac{1}{2} \right] $, and in~\eqref{eq:Plancherel} $
		\mathcal{F}\left\lbrace \cdot \right\rbrace $ denotes the Fourier transform,
	defined as
	\begin{equation}
		\mathcal{F}\left\lbrace x\left(u\right) \right\rbrace = \int_{-\infty}^{\infty}x\left(u\right)e^{-j2\pi\xi u}d\xi.
	\end{equation}
	Line (\ref{eq:Plancherel}) follows from the equivalence of convolution in time
	with multiplication in frequency and Plancherel's theorem. The Fourier
	transforms of the functions in \eqref{eq:Plancherel} equal
	\begin{equation}
		\mathcal{F}\left\lbrace f_{D}\left(u\right)\right\rbrace = \mathcal{F}\left\lbrace \frac{1}{\sqrt{2\pi\left(\frac{\sigma\epsilon}{\Delta}\right)^2}}e^{-\frac{u^{2}}{2\left(\frac{\sigma\epsilon}{\Delta}\right)^2}}\right\rbrace = e^{-2\left(\frac{\pi\sigma\epsilon\xi}{\Delta}\right)^2} \label{eq:fourier_Gaussian},
	\end{equation}
	\begin{equation}
		\mathcal{F}\left\lbrace f_{t}\left(u\right)\right\rbrace = \mathcal{F}\left\lbrace \mathrm{rect}(u) \right\rbrace = \mathrm{sinc}\left(\xi\right) \label{eq:fourier_uniform},
	\end{equation}
	where $ \mathrm{sinc}\left(u\right) = \frac{\sin(\pi u)}{\pi u} $ is the
	normalized sinc function, and
	\begin{align}
		\mathcal{F} & \left\lbrace g\left(u\right) \right\rbrace =
		\mathcal{F}\left\lbrace \mathrm{rect}\left(\frac{u}{2^{k-1}}\right)\ast\sum_{l=-\infty}^{+\infty}\delta\left(u-l2^{k}\right) \right\rbrace                                                   \\
		            & = \mathcal{F}\left\lbrace \mathrm{rect}\left(\frac{u}{2^{k-1}}\right) \right\rbrace \mathcal{F}\left\lbrace \sum_{l=-\infty}^{\infty}\delta\left(u-l2^{k}\right) \right\rbrace \\
		            & = \frac{1}{2}\mathrm{sinc}(2^{k-1}\xi)\sum_{l=-\infty}^{\infty}\delta\left(\xi-\frac{l}{2^{k}}\right) \label{eq:fourier_train}.
	\end{align}

	Substituting \eqref{eq:fourier_Gaussian}, \eqref{eq:fourier_uniform} and
	\eqref{eq:fourier_train} into \eqref{eq:Plancherel}, we get
	\begin{align}
		 & \Pr\left(\whq_{(k)} = q_{(k)}\right) \nonumber                                                                                                                                                                                                    \\
		 & = \frac{1}{2}\int \limits_{-\infty}^{\infty}\! e^{-2\left(\frac{\pi\sigma\epsilon\xi}{\Delta}\right)^2}\!\mathrm{sinc}\left(\xi\right)\mathrm{sinc}(2^{k-1}\xi)\!\!\sum_{l=-\infty}^{\infty}\!\delta\left(\xi-\frac{l}{2^{k}}\right)d\xi\nonumber \\
		 & =\frac{1}{2}\sum_{l=-\infty}^{\infty} e^{-2\left(\frac{\pi\sigma\epsilon l}{\Delta2^{k}}\right)^2}\mathrm{sinc}\left(\frac{l}{2^{k}}\right)\mathrm{sinc}\left(\frac{2^{k-1}l}{2^{k}}\right)                                                       \\
		 & =\frac{1}{2}+\sum_{l=1}^{\infty} e^{-\frac{1}{2}\left(\frac{\pi\sigma\epsilon l}{\Delta2^{k-1}}\right)^2}\mathrm{sinc}\left(\frac{l}{2^{k}}\right)\mathrm{sinc}\left(\frac{l}{2}\right).
	\end{align}

	Finally, the probability of a bit error is
	\begin{align}
		p_k & = \Pr\left(\whq_{(k)} \neq q_{(k)}\right)= 1-\Pr\left(\whq_{(k)} = q_{(k)}\right)                                                                                                        \\
		    & =\frac{1}{2}-\sum_{l=1}^{\infty} e^{-\frac{1}{2}\left(\frac{\pi\sigma\epsilon l}{\Delta2^{k-1}}\right)^2}\mathrm{sinc}\left(\frac{l}{2^{k}}\right)\mathrm{sinc}\left(\frac{l}{2}\right).
	\end{align}
\end{IEEEproof}

\begin{figure}[!t]
	\centering
	\includegraphics[width=0.85\linewidth]{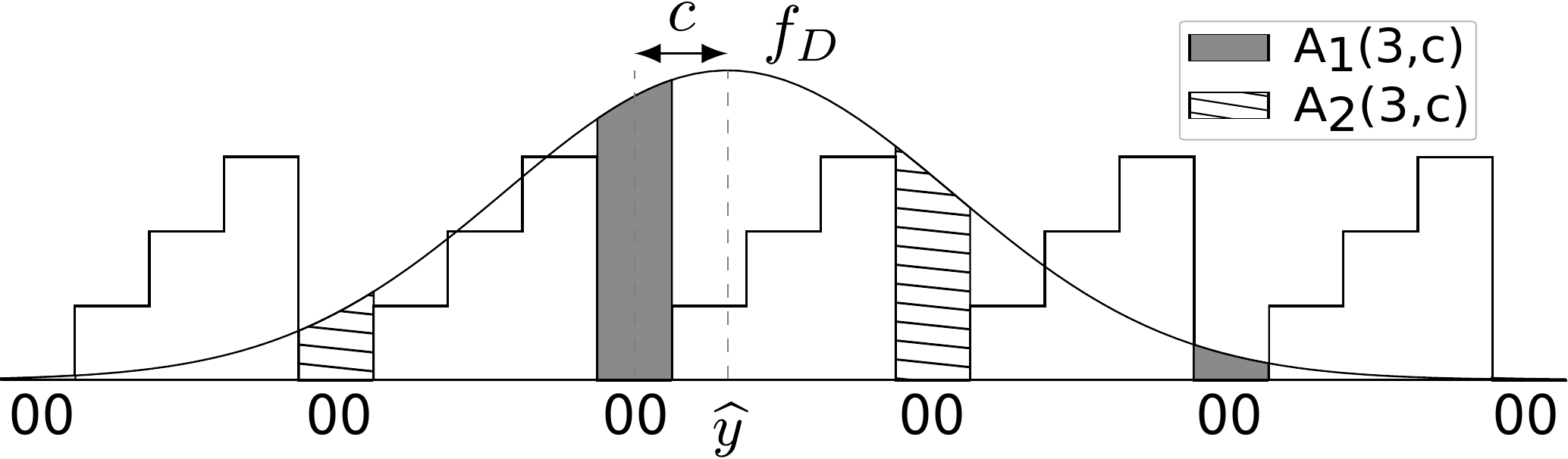}
	\caption{Calculating $ L_{3}(\why) $ by determining the areas $ A_{1}(3,c) $ (gray) and $ A_{2}(3,c) $ (striped).}
	\label{fig:bit_flip_complete}
\end{figure}

\section{Proof of Theorem 2}\label{appx:thm2_prf}
\begin{IEEEproof}
	The proof is similar to the proof of Thm.~\ref{thm:flip_prob} and uses $
		Q_{k-1}(y) $ and $ \sI_{Q_{k-1}(y)} $, as defined, respectively, in
	\eqref{eq:Q_notation} and \eqref{eq:I_notation} in Appendix
	\ref{appx:thm1_prf}. When decoding, the quantity $ c $ denotes the distance
	from $ \why $ to the center of the closest interval consistent with $
		Q_{k-1}(\why) $, that is
	\begin{equation}
		c = \underset{\sI \in \sI_{Q_{k-1}(\why)}}{\mathrm{min}}\left| \why - \frac{\sup(\sI) + \inf(\sI)}{2} \right|.
	\end{equation}
	Fig.~\ref{fig:bit_flip_complete} illustrates the case where the first $ k - 1 =
		2 $ bits are $ 00 $. The quantity $ c $ extends from the predicted measurement
	$ \why $ to the center of the closest $ 00 $ quantization interval.

	Consider the random variable $ D = y - \why $, with pdf $
		f_{D}\left(\cdot\right) $, distributed as $
		\mathcal{N}(0,\left(\frac{\sigma\epsilon}{\Delta}\right)^2) $, as shown in
	Thm.~\ref{thm:flip_prob}. Let $ A_{1}(k, c)$ denote the probability that $ D $
	takes values in the quantization intervals which are consistent with the $ k $
	bits, $ \sI_{Q_{k}(\why)} $, that is
	\begin{equation}
		A_{1}(k, c) = \int_{t \in \sT}f_{D}(t)dt, \label{eq:A1}
	\end{equation}
	where $ \sT = \bigcup_{\sI \in \sI_{Q_{k}(\why)}} \sI$. Similarly, let $
		A_{2}(k, c) $ denote the probability that $ D $ takes values in the
	quantization intervals which are not consistent with $ \sI_{Q_{k}(\why)} $. The
	areas under $ f_{D}(\cdot) $ associated with these two probabilities are shown
	in Fig.~\ref{fig:bit_flip_complete}.

	We estimate the likelihood that the $ k^\mathrm{th} $ bit is flipped as
	\begin{align}
		L_{k} & = \mathrm{Pr}\left(\whq_{(k)} \neq q_{(k)} \mid \why, Q_{k-1}(\why)\right) \\&= \frac{A_{2}(k,c)}{A_{1}(k,c) + A_{2}(k,c)}.
	\end{align}
	We use the $2^k$-periodic rectangular function of unit width
	\begin{equation}
		g(t) = \mathrm{rect}(t)\ast\sum_{l=-\infty}^{+\infty}\delta\left(t-l2^{k}\right),
	\end{equation}
	to express \eqref{eq:A1} as $A_{1}(k, c) = \int_{-\infty}^{+\infty}
		f_{D}(t)g(t - c)dt$, i.e.,
	\begin{align}
		 & A_{1}(k, c) = \int_{-\infty}^{+\infty} \mathcal{F}\left\lbrace f_{D}(t)\right\rbrace \mathcal{F}\left\lbrace g(t-c) \right\rbrace d\xi \label{eqA1}                                                                                                                                   \\
&= \int \limits_{-\infty}^{+\infty}\!\!\!\!\left(e^{-2\left(\frac{\pi\sigma\epsilon\xi}{\Delta}\right)^{\!2}}\right)\!\! \left(\!e^{-i2\pi\xi c}\mathrm{sinc}(\xi)\frac{1}{2^{k}}\!\!\!\sum_{l=-\infty}^{+\infty}\!\!\!\delta\left(\xi-\frac{l}{2^{k}}\right)\!\right)\! d\xi\nonumber \\
		 & = \frac{1}{2^{k}} \int \limits_{-\infty}^{+\infty}\! e^{-2\left(\frac{\pi\sigma\epsilon\xi}{\Delta}\right)^2}\cos(2\pi\xi c) \mathrm{sinc}(\xi)\!\!\sum_{k=-\infty}^{+\infty}\!\!\!\delta\left(\xi-\frac{k}{2^{k}}\right)\!d\xi\nonumber                                              \\
		 & = \frac{1}{2^{k}}\!\Biggl(1 + 2 \!\sum_{l=1}^{+\infty}\! e^{-\frac{1}{2}\!\left(\frac{\pi\sigma\epsilon l}{2^{k-1}\Delta}\right)^{\!2}}\!\!\cos\!\left(\!\frac{\pi ck}{2^{k-1}}\!\right)\!\mathrm{sinc}\!\left(\!\frac{l}{2^{k}}\!\right)\!\!\Biggr).
	\end{align}
Similarly, we note that inconsistent quantization intervals are shifted by $ 2^{k-1} $ with respect to consistent intervals, and hence
	\begin{align}
		 & A_{2}(k, c) = A_{1}(k, 2^{k-1} - c) \nonumber                                                             \\
&= \frac{1}{2^{k}}\!\Biggl(\!1 + 2\!\sum_{l=1}^{+\infty}\!e^{-\frac{1}{2}\!\left(\!\frac{\pi\sigma\epsilon l}{2^{k-1}\Delta}\!\right)^2}\!\!\!\cos\!\left(\!\!\frac{\pi l\!\left(2^{k-1}-c\right)}{2^{k-1}}\!\!\right)\!\!\mathrm{sinc}\!\left(\!\frac{l}{2^{k}}\!\right)\!\!\Biggr).
	\end{align}
\end{IEEEproof}
\begin{IEEEbiography}[{\includegraphics[width=1in,height=1.25in,clip,keepaspectratio]{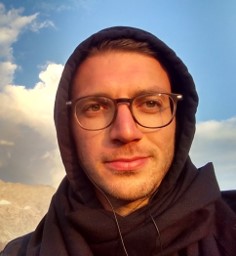}}]{Maxim
        Goukhshtein} (GS'16) is currently pursuing a Ph.D. degree in electrical and
    computer engineering at the University of Toronto, Toronto, ON, Canada. He
    received the B.Eng. degree in electrical engineering from McGill
    University, Montreal, QC, Canada in 2015, and the M.A.Sc. degree in
    electrical and computer engineering from the University of Toronto in 2017.

    He completed engineering internships at National Instruments, Austin, TX,
    USA and Ericsson, Ottawa, ON, Canada, and was a Research Intern at the
    Mitsubishi Electric Research Laboratories (MERL), Cambridge, MA, USA. His
    research interests lie at the intersection of information theory, coding
    theory and signal processing. Most recently he has been working on
    probabilistic shaping and source resolvability. He was previously supported
    by the Queen Elizabeth II Graduate Scholarship and the Ontario Graduate
    Scholarship, and currently by the NSERC Postgraduate Scholarship.
\end{IEEEbiography}

\begin{IEEEbiography}[{\includegraphics[width=1in,height=1.25in,clip,keepaspectratio]{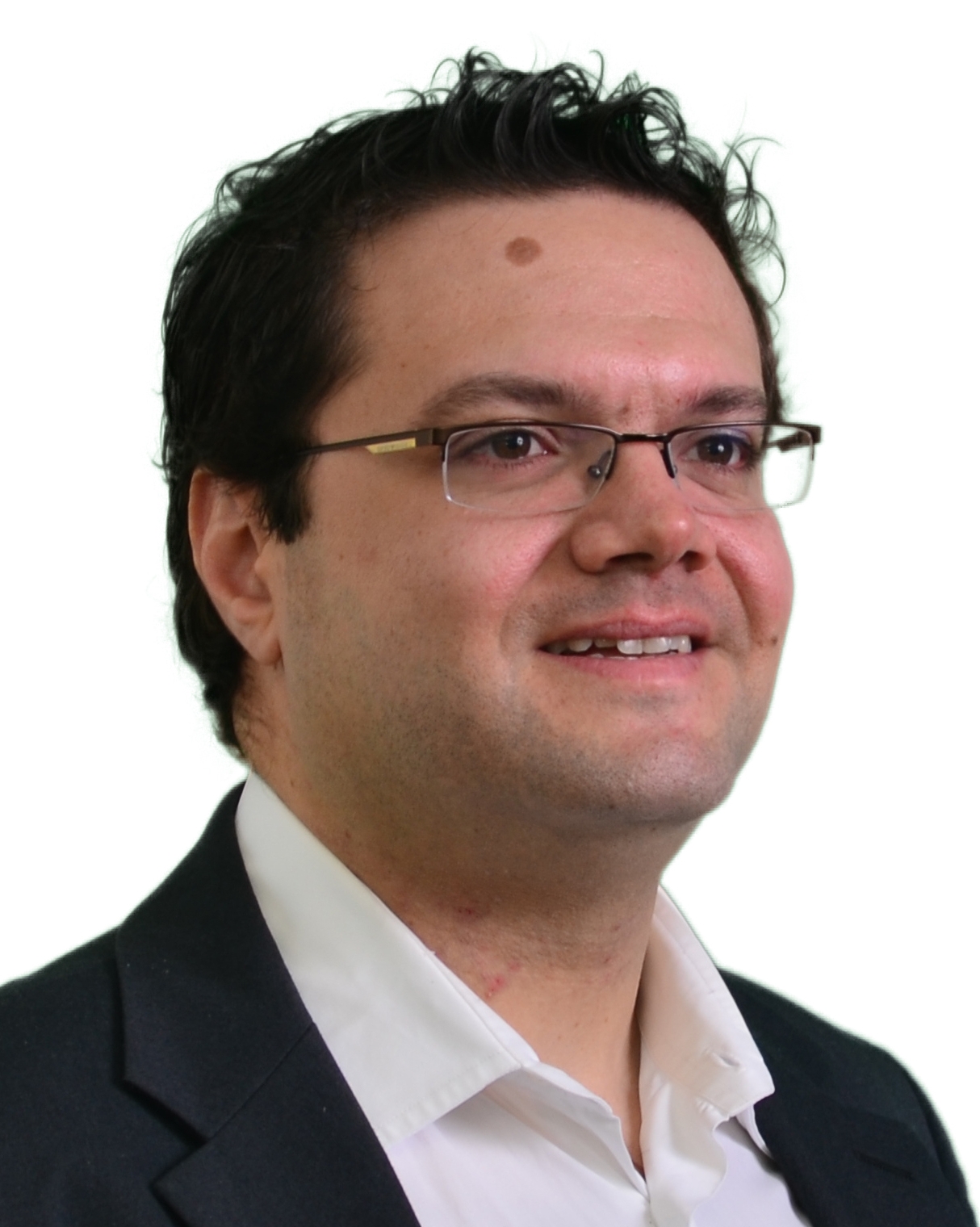}}]{Petros
        T.\ Boufounos} (S'02-M'06-SM'13) is a Senior Principal Research
    Scientist and the Computational Sensing Team Leader at Mitsubishi
    Electric Research Laboratories (MERL), and a visiting scholar at the
    Rice University Electrical and Computer Engineering department.
    Dr.~Boufounos completed his undergraduate and graduate studies at MIT.
    He received the S.B.~degree in Economics in 2000, the S.B.\ and M.Eng.\
    degrees in Electrical Engineering and Computer Science (EECS) in 2002,
    and the Sc.D.\ degree in EECS in 2006. Between September 2006 and
    December 2008, he was a postdoctoral associate with the Digital Signal
    Processing Group at Rice University. He joined MERL in January 2009.

    Dr.~Boufounos' immediate research focus includes signal acquisition and
    processing, inverse problems, frame theory, quantization and data
    representations. He is also interested into how signal acquisition
    interacts with other fields that use sensing extensively, such as machine
    learning, robotics and dynamical system theory. Dr. Boufounos has served as
    an Associate Editor and a Senior Area Editor at IEEE Signal Processing
    Letters, has been part of the SigPort editorial board, and is currently a
    member of the IEEE Signal Processing Society Theory and Methods technical
    committee and an Associate Editor at IEEE Transactions on Computational
    Imaging.
\end{IEEEbiography}

\newpage

\begin{IEEEbiography}[{\includegraphics[width=1in,height=1.25in,clip,keepaspectratio]{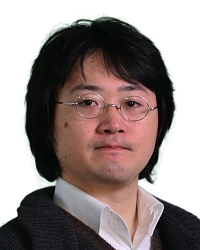}}]{Toshiaki
        Koike-Akino} (M’05–SM’11) received the B.S. degree in electrical and
    electronics engineering, M.S. and Ph.D. degrees in communications and
    computer engineering from Kyoto University, Kyoto, Japan, in 2002,
    2003, and 2005, respectively. During 2006–2010 he was a Postdoctoral
    Researcher at Harvard University, and joined MERL, Cambridge, MA, USA,
    in 2010. His research interests include signal processing for data
    communications and sensing. He received the YRP Encouragement Award
    2005, the 21st TELECOM System Technology Award, the 2008 Ericsson Young
    Scientist Award, the IEEE GLOBECOM’08 Best Paper Award, the 24th
    TELECOM System Technology Encouragement Award, and the IEEE GLOBECOM’09
    Best Paper Award.
\end{IEEEbiography}

\begin{IEEEbiography}[{\includegraphics[width=1in,height=1.25in,clip,keepaspectratio]{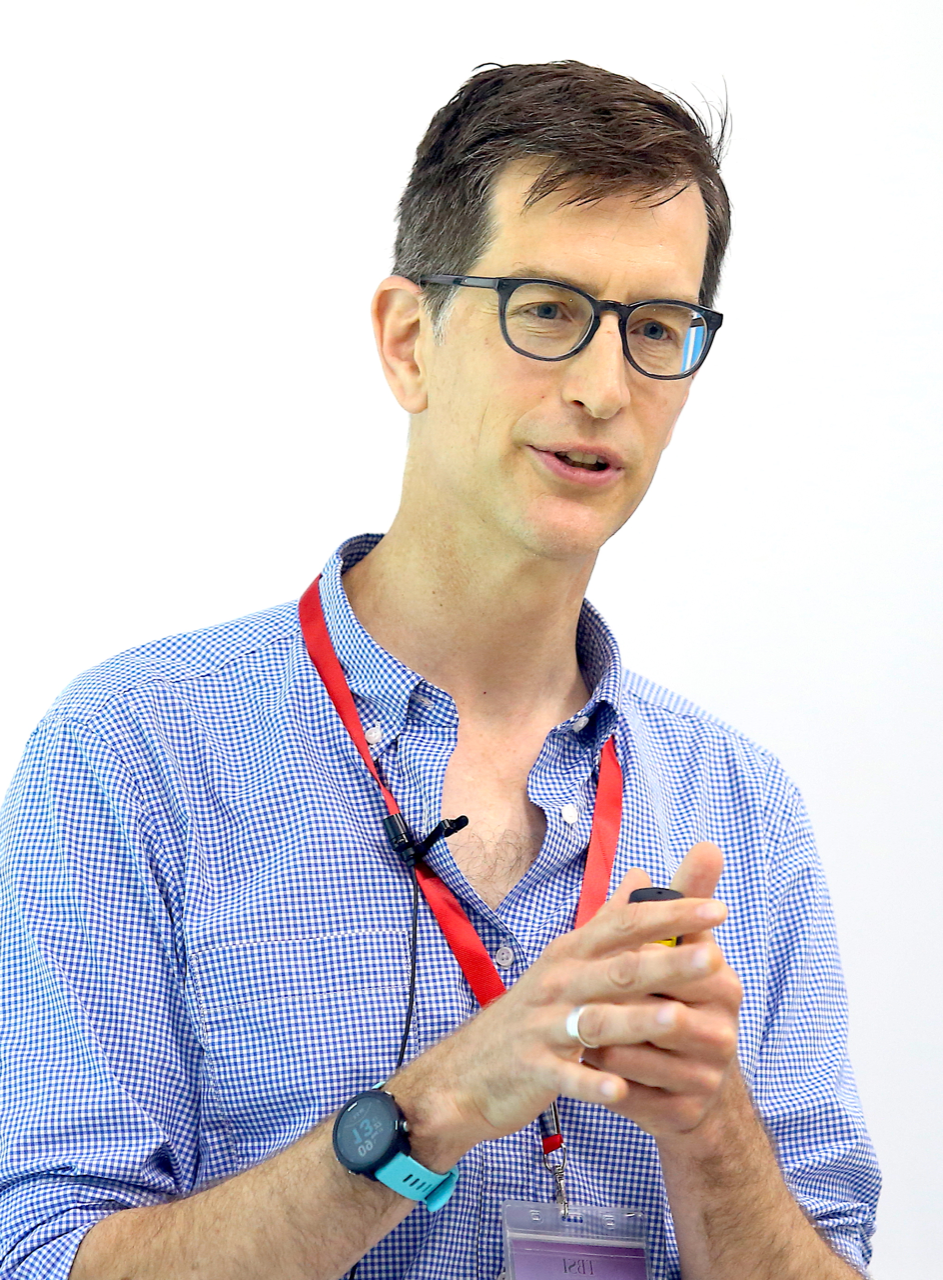}}]{Stark
        C.\ Draper } (S’99-M’03-SM’15) is a Professor of Electrical and
    Computer Engineering at the University of Toronto (UofT) and was an
    Associate Professor at the University of Wisconsin, Madison. As a
    research scientist he has worked at the Mitsubishi Electric Research
    Labs (MERL), Disney’s Boston Research Lab, Arraycomm Inc., the C. S.
    Draper Laboratory, and Ktaadn Inc. He completed postdocs at the
    University of Toronto and at the University of California, Berkeley. He
    received the M.S. and Ph.D. degrees from the Massachusetts Institute of
    Technology (MIT), and the B.S. and B.A. degrees in Electrical
    Engineering and in History from Stanford University. His research
    interests include information theory, optimization, error-correction
    coding, security, and the application of tools and perspectives from
    these fields in communications, computing, and learning.

    Prof. Draper has received the NSERC Discovery Award, the NSF CAREER Award,
    the 2010 MERL President’s Award, and teaching awards from the University of
    Toronto, the University of Wisconsin, and MIT. He received an Intel
    Graduate Fellowship, Stanford’s Frederick E. Terman Engineering Scholastic
    Award, and a U.S. State Department Fulbright Fellowship. He spent the
    2019-20 academic year on sabbatical at the Chinese University of Hong Kong,
    Shenzhen, and visiting the Canada-France-Hawaii Telescope (CFHT) in Hawaii,
    USA.  He chairs the Machine Intelligence major at UofT and serves on the
    IEEE Information Theory Society Board of Governors.
\end{IEEEbiography}

\end{document}